\documentclass[fleqn,twoside]{article}
\usepackage{espcrc2}

\usepackage{graphicx}
\newcommand{\AmS}{{\protect\the\textfont2
  A\kern-.1667em\lower.5ex\hbox{M}\kern-.125emS}}

\title{Hadrons and Chiral Symmetry}
\author{Thomas D. Cohen\address[MCSD]{Department of Physics,
        University of Maryland \\
        College Park, Maryland 20742, USA}}

\begin{document}

\begin{abstract}
These lectures discuss the question of whether a key feature is seen in hadron spectroscopy---the near degeneracy of hadrons with different parity and/or spin.  It has been conjectured that this is due to an effective restoration of chiral symmetry.  The conjecture is that while these states are, of course, in the symmetry-broken (Nambu-Goldstone) phase, as one goes higher in the spectrum the states become progressively less sensitive to the dynamics of chiral symmetry breaking.  These lectures discuss the current status of this conjecture.  The motivations for the conjecture are discussed, as is evidence---both theoretical and experimental---in its favor.  Possible alternative explanations for the data are also discussed.
\vspace{1pc}
\end{abstract}


\maketitle

\section{Introduction}

Let me state at the outset that I consider the conjecture of effective chiral restoration to be both intriguing and highly speculative.  These lectures will I hope convey both of these aspects.  These lectures are rather informal and personal and are intended to introduce the basic issues and to give my reading of  my current take on the state of the art.  They are {\it not} intended as a complete technical review and do not discuss fully all of the literature in the field.  Moreover, as these lectures are for a school, rather than a workshop, I will review a number of subjects which are an important part of the tool box for people working in this field but are not necessarily taught in standard courses.  For a more  complete and technical discussion of the issue of effective chiral restoration, one should consult Lenya Glozman's recent review article\cite{GlozmanReview}.  Glozman has been a driving force behind this idea and has thought deeply about the subject for many years.

\subsection{An odyssey}
Let me begin with a brief description of my personal odyssey of how I became involved in this field.  I start this way for a couple of reasons.  The style of doing theory in this field is rather different than in most.  It is driven largely by data and rather qualitative reasoning.  It is certainly different from the other lectures at this school; it is also quite different from the style of most of the other research I have been involved with over my career.  It is this peculiar nature of the field that makes me both intrigued by and simultaneously somewhat skeptical of the conjecture.

I originally became acquainted with the idea in the summer of 2000.  At the time there was a well-recognized issue in hadronic physics---the problem of parity doubling in the baryon spectrum.  While the origin of this phenomena was obscure, it was clear to anybody even casually looking at the data that many excited nucleon states had ``parity partners''---states of the same spin and opposite parity which were {\it nearly} degenerate.   Examples of these include the $N^*(1675)$  and  $N^*(1680)$ (positive and negative parity spin 5/2 resonances), $N^*(1720)$  and  $N^*(1700)$ (positive and negative parity spin 3/2 resonances) and $N^*(2220)$  and  $N^*(2250)$ (positive and negative parity spin 9/2 resonances).  Lenya Glozman\cite{G1}had a novel explanation for these---namely that they were the result of chiral restoration.  I learned of this explanation from Glozman at a workshop at Lake Bled in Slovena.  At the time, I must confess that I did not think much of it.

My worry with the original formulation of this explanation was two-fold.  In the first place it was set in the context of a quark model based on Goldstone boson exchange between the quarks\cite{GR}.  Now, I have never been a big fan of quark models in general since their connection to QCD has always been obscure.  Moreover, this particular form of the quark model was highly unconventional in the context of these models. A second problem was that the phenomenon was described as a ``phase transition''\cite{G1}.  However, this was clearly not a phase transition in anything resembling the usual meaning of the word.  It does not describe anything thermodynamic, nor is there any discontinuity.  In retrospect, I was clearly a bit hasty in forming a negative view of the idea.  (I will note that I was not alone in this).  As it happens, the description in terms of a quark model is completely inessential to the idea, and the description as a ``phase transition''  was really nothing more than an infelicitous use of language.

My next interaction with the notion of effective chiral restoration  came at a rather surreal time.  The hadronic community in the US had organized a workshop to produce a white paper on priorities in  hadronic physics.  The workshop was held in Duck, North Carolina in early November 2000.  The surreal aspect was that it occurred over election day in the US---an election in which the news media first announced that the great state of Florida had been won by Al Gore and then, in effect, said ``oops---within error bars its a tie.''  During this truly odd period, major outstanding issues of hadronic physics were discussed.  One of these was parity doubling in the nucleon spectrum.  Somebody remarked that ``Nobody has an explanation for this.'',  to which I called attention to Glozman's ideas---taking care neither to endorse nor reject it.  Bob Jaffe then made a profound comment\cite{Duck}:  He argued that Glozman must be wrong since if he were correct one would expect chiral multiplets rather than parity doublets.  He went on to suggest an effect associated with the $U_A(1)$ anomalous current.

For reasons which I will describe later, I felt that Jaffe's suggestions about the $U_A(1)$ anomalous current were off target.  However, his critique of Glozman's argument seemed to me to be trenchant.  I communicated this argument to Glozman.  He then asked me what the multiplets would look like assuming that Jaffe was correct.  As it happened, in conjunction with Xiangdong Ji, I had previously worked out chiral multiplets with baryon quantum numbers\cite{CJ}.  This work with Ji was intended for application of QCD at finite temperature where chiral symmetry could indeed be restored in the sense of a phase transition.  Thus, I was able to send these to Glozman rather quickly.  Almost immediately I received a reply from him  to the effect he had looked in the particle data book and the patterns seen in the data are in accord with the multiplets enumerated in ref.~\cite{CJ}.  This came with a suggestion---or perhaps it was a command---``lets write a paper!''.

Thus began my foray into this field.  There were two main aspects of this which will be described in detail in these lectures.  One of these was phenomenological.  The central question there is to what extent does the data support the conjecture?  My current read on this is that the data is suggestive but not compelling.  The other aspect is theoretical.  Ideally the central theory question would be how to obtain the phenomenology directly from QCD.  However, there is no obvious path as to how to do this.  Thus, we are compelled to consider a more modest goal: namely to establish that the conjecture is not {\it inconsistent} with what we know about QCD.  In effect, this goal is simply to demonstrate that the idea is not crazy.  However, even this modest goal is non-trivial:  the idea of effective chiral restoration in the spectrum strikes many people to be crazy in a number of ways.  My current read on this front is that the conjecture is not crazy {\it a priori}.  There is nothing {\it known} about QCD that is inconsistent with the conjecture.  Whether the conjecture is ultimately the explanation for the data, however, remains an important open question.
It should be clear from this Odyssey that I remain  both intrigued by the conjecture of effective restoration but also skeptical that it is correct.  I would note that despite this skepticism, I do believe that the topic is completely appropriate for a winter school dedicated to open challenges in QCD.   Whatever the ultimate truth of the conjecture, the issue of how to treat highly excited hadrons is clearly an open challenge in QCD.

\subsection{Highly excited hadrons}

The Particle Data Book\cite{PDG} contains an immense amount of information.  Almost all of it pertains to hadronic resonances.  The amount of information about these states vastly exceeds the information about stable hadrons.  Thus, it is probably fair to say that {\it anything}  we can learn about these states which directly connects to QCD is of some importance.  Unfortunately, there is no simple way   to attack this problem from QCD.  In fact, the usual tool for the study of highly excited hadrons is a model; some variant of the constituent quark model is typically employed.  At first sight this seems sensible---the Particle Data Book lists masses for the hadrons and the constituent quark model produces masses which can then be compared with this data.

There are two problems with such an approach.  The first was alluded to earlier: the connection of the constituent quark model and QCD remains obscure.  Thus, it is by no means clear {\it what} one learns from a quark model fit.  The second issue is more problematic and concerns the nature of the spectroscopic data contained in the Particle Data Book.  It was famously said by Voltaire that the Holy Roman Empire was neither Holy, nor Roman, nor an Empire.  Something similar could be said about the Particle Data Book: it is largely neither about particles, nor composed of data, nor a book.  Rather, in it is a web site \cite{PDG2} largely composed of model-dependent fits about resonances in scattering processes.  In a very important sense, the quark model---in its simple incarnation---lives in a different world than the data.  The quark model yields masses while the actual data is for differential cross sections in scattering processes.  If the resonance is isolated and narrow one {\it might} hope that some essential feature of it may be captured in a model yielding more-or-less clearly defined masses.  But this is clearly problematic for the case of highly excited states.  Exactly how one models the backgrounds can alter the property of the resonances.  This intrinsic ambiguity in the masses should raise a caution flag to any approach which matches a theoretical approach directly to the ``data'' of the hadron masses.

\subsection{Resonances as scattering amplitudes}

The difficulties associated with the fact that ambiguities in hadron masses can in principle be avoided.  Rather than focusing on the hadrons {\it per se} with their attendant ambiguities, one can attempt to describe theoretically the physical observables themselves.  These observables are associated with scattering processes of various sorts and are encoded quantum mechanically in scattering amplitudes.  These scattering amplitudes are, at least in principle, directly calculable from QCD.

In practice, however, the only known method to compute scattering amplitudes directly in QCD is via the L\"uscher method\cite{lat}.  The trick here is to use the fact that a lattice is necessarily finite in spatial extent and the imposition of finite size boundary conditions renders the spectrum discrete.  The spacing of the energy levels and their variation of this spacing with the size of the lattice is connected in a known way with the phase shifts of the scattering process at that energy for the continuum  of the system, {\it provided the size of the system is sufficiently larger than the range of the interactions so that spurious finite size effects are negligible}.  While this approach may be a viable approach to calculations of low energy scattering observables such as scattering lengths, it is intrinsically very difficult to use to trace out scattering amplitudes for resonances. The essential problem is that to map out the amplitudes as a function of energy through the resonance one needs to have {\it many} energy eigenvalues through the resonance region.  This can be achieved at least in principle by making the lattice size very large.  While this is also valuable from the perspective of limiting spurious finite size effects it raises two fundamental difficulties. The first is simply that large sizes are intrinsically expensive numerically.  The second is that signal to noise drops exponentially with each level so that accurately computing the energy of many levels is truly daunting.  Given this problem, the computation of scattering amplitudes for highly excited  resonances is likely to be far off in the future.

This raises a key question---is it possible to learn {\it anything} about highly excited resonances from QCD in a model-independent way?  The answer is yes---at least for baryons in the large $N_c$ limit.  As we will see, the amount we can learn about spectroscopy in a model-independent way is ultimately rather limited.  Nevertheless, the fact that one can learn anything makes this worth pursuing.  Moreover, discussing this approach is a useful way to introduce the idea of large $N_c$ QCD and the $1/N_c$ expansion which will play a central role in these lectures.

\subsection{Large $N_c$ QCD}

The fundamental problem with QCD at low momenta is the absence of a natural expansion parameter.  However, within a year of the formulation of QCD, `t~Hooft suggested that one can generalize QCD from 3 colors to $N_c$ colors and then use $1/N_c$ as an expansion parameter\cite{HOOFT}.  The notion underlying this suggestion is that the world with $N_c=3$ is qualitatively similar to the large $N_c$ world.  Thus physical observables will be taken to be a series where the leading term is its large $N_c$ value and subsequent terms are powers in $1/N_c$.

As it happens, except in 1+1 dimension\cite{HOOFT1}, we do not know how to solve QCD at large $N_c$.  However, the $1/N_c$ expansion remains a valuable tool.  While one cannot solve the theory even at leading order, one can deduce how various observables scale with $N_c$.  The basic strategy developed by `t Hooft, is to focus on the color flow within a Feynman diagram.  To aid this a clever ``double line'' notation was introduced for gluons.  Using these diagrams it quickly becomes apparent that a  consistent large $N_c$ limit requires the following simultaneous limits:
\begin{equation}
N_c \rightarrow \infty \; \; \; \; \; \;g \rightarrow 0  \; \; \; \; \; \; \frac{g}{\sqrt{N_c}} \; {\rm fixed} \; .
\end{equation}
With this limiting process and the double line notation, it is easy to show that the leading order diagrams are planar, that each non-planar gluon costs two powers of $1/N_c$, that each quark loop costs one power of $1/N_c$; and that the leading diagrams containing a quark loop have the loop bounding the diagram.

These have important implications for correlation functions for currents with the quantum numbers of mesons.  From these it is possible to deduce the $N_c$ scaling of mesonic properties:
\begin{eqnarray}
m_{\rm meson} & \sim & N_c^0  \nonumber \\
\Gamma_{3-\rm meson}  & \sim & N_c^{-1/2} \nonumber \\
\Gamma_{n-\rm meson}  & \sim & N_c^{1-n/2} \; ,
\label{meson-scaling}
\end{eqnarray}
where $\Gamma$ represents a multi-meson vertex.  An important consequence of these rules is that at large $N_c$ mesons become weakly interacting. Of particular importance for these lectures is the fact that these rules imply that at large $N_c$ the mesons become long-lived, {\it i.e.}, the resonances become narrow.  This is significant in that in a large $N_c$ world, ambiguities about meson masses disappear.

Witten extended large $N_c$ analysis to baryons\cite{WittenN}.  The basic approach was to show that  in a large $N_c$ world baryons could be described self-consistently in a mean-field picture (at least for heavy quarks) from which large $N_c$ scaling rules could be inferred:
\begin{eqnarray}
m_{\rm baryon} & \sim & N_c  \nonumber \\
g_{\rm meson-baryon}  & \sim & N_c^{1/2} \nonumber \\
g_{\rm 2 meson-baryon}  & \sim & N_c^{0} \; ,
\label{baryon-scaling}
\end{eqnarray}
where $g$ is the coupling strength.  These rules imply  that baryons are heavy at large $N_c$.   These scaling results also suggest an interesting tension.  On the one hand mesons are strongly coupled to baryons; the coupling scales as $N_c^{1/2}$.  On the other hand, the baryon-meson scattering amplitude (as encoded by $g_{\rm 2 meson-baryon}$) is independent of $N_c$.

There are important phenomenological results of these generic $N_c$ scaling rules.  For example the OZI rule is naturally explained: it becomes exact at large $N_c$.  Similarly, the phenomenological fact that mesons tend to decay dominantly into two mesons---which themselves tend to decay into mesons (if possible).  Again this becomes exact at large $N_c$.  The large $N_c$ world can be shown not to have exotic two-quark--two-antiquark states and this may explain the absence of such states in nature.  Other aspects of the large $N_c$ world include the existence of an infinite number of mesons with any quantum number (which may partially explain the fact that there are several in the real world) and, similarly, that there are an infinite number of glueballs (none of which have been seen definitively in our $N_c=3$ world); moreover, at large $N_c$, glueballs and mesons are unmixed.   The large $N_c$ world also requires the existence of exotic ``hybrid'' mesons with quantum numbers which cannot be written as pure quark-antiquark states\cite{hybrid}.  The Regge picture may be justified in part in the large $N_c$ world due to the domination of glueball and meson tree graphs in the effective theory.

\subsection{Spin-flavor symmetry}

All of the phenomenological large $N_c$ results described above are qualitative. It turns out that there are some quantitative (or at least semi-quantitative) results as well.  The scaling rules in Eqs.~(\ref{meson-scaling}) and (\ref{baryon-scaling}) are generic.  They make no reference to spin and flavor.  For baryons, spin and flavor play a critical role.  The key idea is large $N_c$ consistency conditions\cite{GS1,GS2,DM,Jenk,DJM1,DJM2}.
Ultimately, these conditions imply the existence of an emergent spin-flavor symmetry at large $N_c$; it is a contracted
SU(4) symmetry (assuming two flavors).  This symmetry requires a tower of degenerate states with $I=J$ (where I is the isospin).  In the real world of $N_c=3$ the nucleon and the $\Delta$ correspond to states in this tower---higher spins and isospins only exist for $N_c > 3$.  At finite $N_c$, the $\Delta$  is not degenerate with the nucleon but their splitting can be shown to scale as $1/N_c$.

The contracted SU(4) symmetry is also an important implication for the coupling between states in the tower. There are four basic types of couplings: unity, a spin operator, an isospin operator, and a spin-isopsin operator $X$.  These operators are generators of an algebra.  It is given by:
\begin{eqnarray}
\left [J_j,J_k \right ] & =  & i \, \epsilon_{jkl} J_l \nonumber \\
\left[ I_a,I_b \right] & = &i \, \epsilon_{abc} I_c \nonumber \\
\left[ J_j,X_{k,a}\right ]& = &i \, \epsilon_{jkl} X_{la}  \nonumber \\
\left[ I_a,X_{j,b} \right]& = &i \, \epsilon_{abc} X_{jc} \nonumber \\
\left[ X_{ja},X_{kb}\right ]&=& 0 \; .
\label{contracted}
\end{eqnarray}

The fact that the last commutator vanishes makes this a contracted symmetry. In deriving this symmetry, the central issue is the fact that baryon-meson coupling is order $N_c^{1/2}$.  Thus, the Born and cross-Born graphs in pion-nucleon scattering are each of order $N_c^1$ but unitarity implies that the scattering amplitude is of order unity.  This implies cancelations must occur.  This cancelation is summarized in the vanishing commutator in Eq.~(\ref{contracted}).  The tower of baryon states with $I=J$ follows if the contracted symmetry is in its simplest nontrivial representation.

\subsection{Scattering amplitude relations \label{SAR}}

\begin{figure}
\centering
\includegraphics[width=3.in]{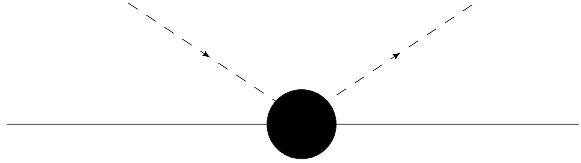}
\caption{A $\pi$-N scattering process.  The grey blob representing the process can be viewed as an operator acting in the baryon space. }
 \label{pi_N}
\end{figure}

This contracted symmetry fixes relations among static observables.  However, as described in a series of papers~\cite{CL1,CL2,CL3,CL4,CL5,CL6,CL7,CL8,CL9}, it also fixes relations among scattering amplitudes.  The easiest way to see this is to focus on a very simple process: pion-nucleon scattering.  Now from the perspective of a nucleon the one can visualize the process as in Fig.~\ref{pi_N}.  The grey blob representing the scattering is an operator in the space of baryons and hence is describable in terms of the contracted symmetry.  From the perspective of the symmetry all operators which contribute in leading order have $I=J$ in the sense of the Wigner-Eckhart theorem in space and isospace.  That is, they can be scalar-isoscalar or vector-isovector operators, but scalar-isovector or vector-isoscalar operators are down by one power $1/N_c$.  From the perspective of scattering, these operators correspond to t-channel amplitudes.  Clearly, the large $N_c$ limit constrains the possible scattering amplitudes since the most general amplitude is not restricted to $I=J$ contributions.

The natural language to describe resonances is not in the t-channel but in the s-channel.  One can label the amplitude by initial (final) orbital angular momentum, $L$ ($L'$), and the s-channel spin and isospin $J$ and $I$, respectively.  Imposing $I=J$ in the t-channel and using standard identities enables one to deduce that the scattering amplitude can be written as
\begin{eqnarray}
&~&S_{LL^\prime  IJ}   =
3 \sum_K  (2K+1) \nonumber \\
&\times & \left\{ \begin{array}{ccc} K &
I & J\\ 1 & L^\prime & 1 \end{array} \right\} \left\{
\begin{array}{ccc} K & I & J \\ 1 & L & 1 \end{array} \right\}
s_{KL^\prime L} , \label{MPeqn1}
\end{eqnarray}
where the $s_{KL^\prime L}$ are ``reduced amplitudes'' which are functions of the external momenta.  Since there are more amplitudes than reduced amplitudes, various amplitudes are equal to  fixed linear combinations of other amplitudes at large $N_c$.  Thus, while one cannot directly compute any scattering amplitude, having measured some amplitudes one can then predict others. These predictions are correct up to $1/N_c$ corrections; typically they work fairly well. This approach can be easily extended in many directions including to arbitrary mesons, to final $\Delta$ states, to photoproduction amplitudes, to next-to-leading order in the $1/N_c$ expansion and to SU(3) flavor.

In many ways this approach gives significant insights into the physics of excited baryon states.  Moreover, the approach has real predictive power at the level of amplitudes. In principle, it makes important predictions for relations among baryon resonances.  Before discussing this in detail, it is worth recalling that while meson resonances generically become narrow at large $N_c$ due to the small couplings, this does not happen for baryons: a typical baryonic resonance has a width of order $N_c^0$.  Nothing about large $N_c$ scaling implies that resonances should be narrow---or even exist at all.  However, assume that a resonance narrow enough to be detected {\it does} exist.  Such a resonance can be interpreted as a pole in the scattering amplitude at complex momentum.   The large $N_c$ scaling rules in Eq.~(\ref{MPeqn1}) give a physical amplitude as a sum of contributions of reduced matrix elements times group theoretic factors.  Thus if there is a pole in such an amplitude there must be a pole in the reduced amplitude.  However, since the same reduced amplitude contributes to multiple physical amplitudes, there must be poles {\it at the same place} in multiple amplitudes.  To the extent that one can determine the mass and width of a resonance reflects the position of the pole, this means that in a large $N_c$ world baryon resonances would come in  multiplets with degenerate masses and widths.  These multiplets would be labeled by the $K$ quantum number.

Unfortunately, this striking prediction about degenerate multiplets of baryon resonances in a large $N_c$ world is problematic for the world of $N_c=3$. This can be seen by a cursory look at the masses and widths extracted from the scattering data.  The critical problem is that the splittings within a given multiplet due to $1/N_c$ effects is as large as the splitting between multiplets.   This suggests that as far as this key feature is concerned, $N_c=3$ is simply too small a number for the $1/N_c$ expansion to be useful.

\subsection{A Baconian approach}

Given the intrinsic limitations of models, the difficulties with $1/N_c$ expansion for excited baryons and the present intractability of lattice QCD for highly excited states, one might ask how we can hope to learn {\it anything} fundamental about excited hadrons.  One possible approach to this is old as science itself---Francis Bacon's notion that one learns about nature by looking for patterns in the empirical data and then form an hypothesis about them which can subsequently be tested.

Now this Baconian approach has an intrinsic limitation in the case of excited hadrons in that the actual data---scattering data---is quite voluminous.  To have any reasonable hope of finding patterns, it is probably necessary to use massaged data---extractions of hadronic masses and, perhaps, widths from the data.   As noted earlier, this introduces some ambiguities: such extractions necessarily have some model dependence.  The hope is that these ambiguities are small enough so that real lessons can be drawn from the data.

\subsection{Parity doublets \label{PD}}

\begin{figure*}
\centering
\includegraphics[width=4.5in, angle=-90]{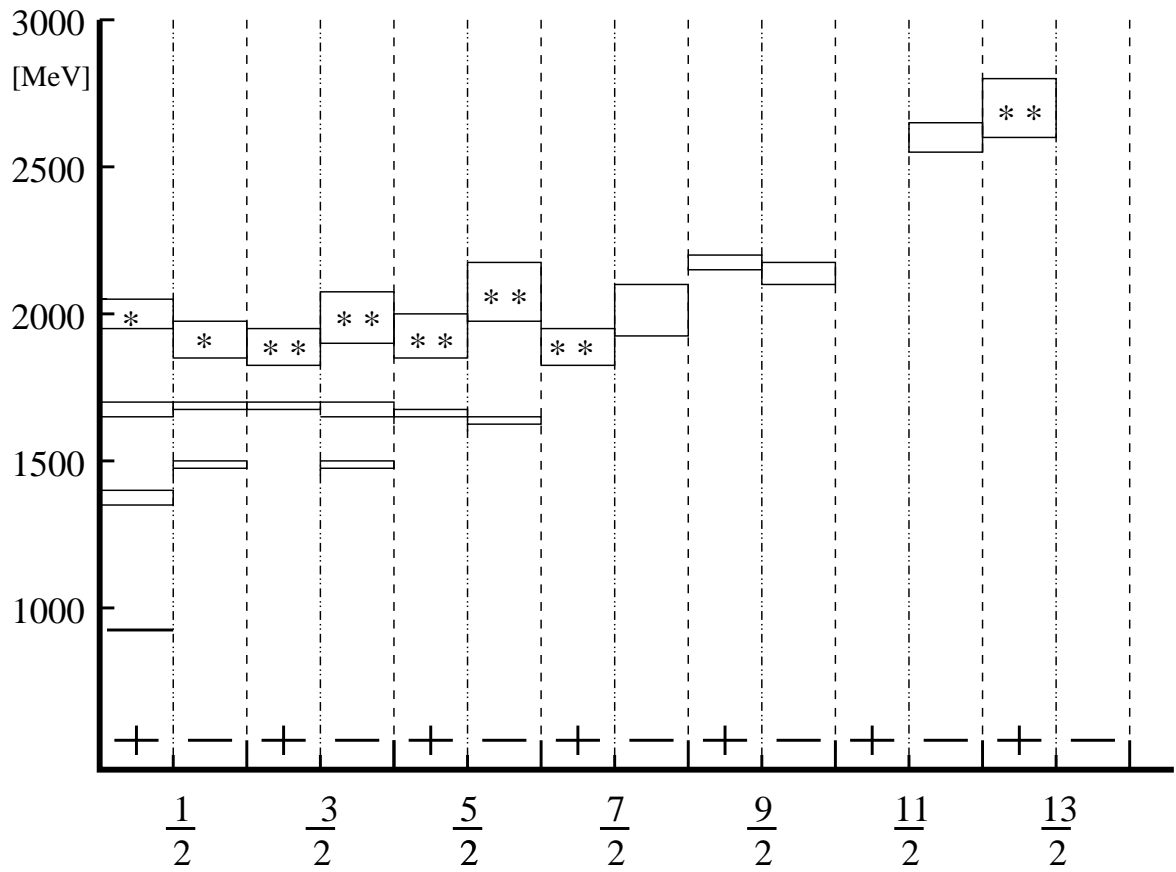}
\label{ns}
\caption{Nucleon resonances.  Some of these states are
not well established.  They are ``one star'' or ``two star'' states in the classification scheme of the Particle Data Group and  marked accordingly on the figure.  This figure is from ref.~\cite{GlozmanReview}}
\end{figure*}

It was realized long ago that ``parity doublets'' were a common feature in the baryon spectrum.  Parity doublets refer to two baryon masses with identical quantum numbers except for parity which are nearly degenerate. A classic case of these are the positive parity spin 5/2 $N^*$(1675) and the negative parity spin 5/2 $N^*$(1680).  The question is how common are these.  In fact, they are quite common---one could indeed argue that they are ubiquitous.  The question one might ask is whether they really {\it are} ubiquitous in the sense that {\it all} high-lying baryon resonances fall into such doublets. A quick look at Fig.~\ref{ns} suggests that they might be.  Moreover, the pattern is similar with $\Delta$ excitations.  If it is really the case there might be a deep explanation for these doublets.

Now before going on, the question one might ask is how compelling is the data that {\it all} high-lying nucleons are in parity doublets.  I would suggest that the data, while not compelling, is certainly suggestive.   It is hardly surprising that it is not compelling: it is very difficult to get ``smoking gun'' quality evidence simply by looking at the extracted masses.

\begin{figure*}
\centering
\includegraphics[width=6.0in]{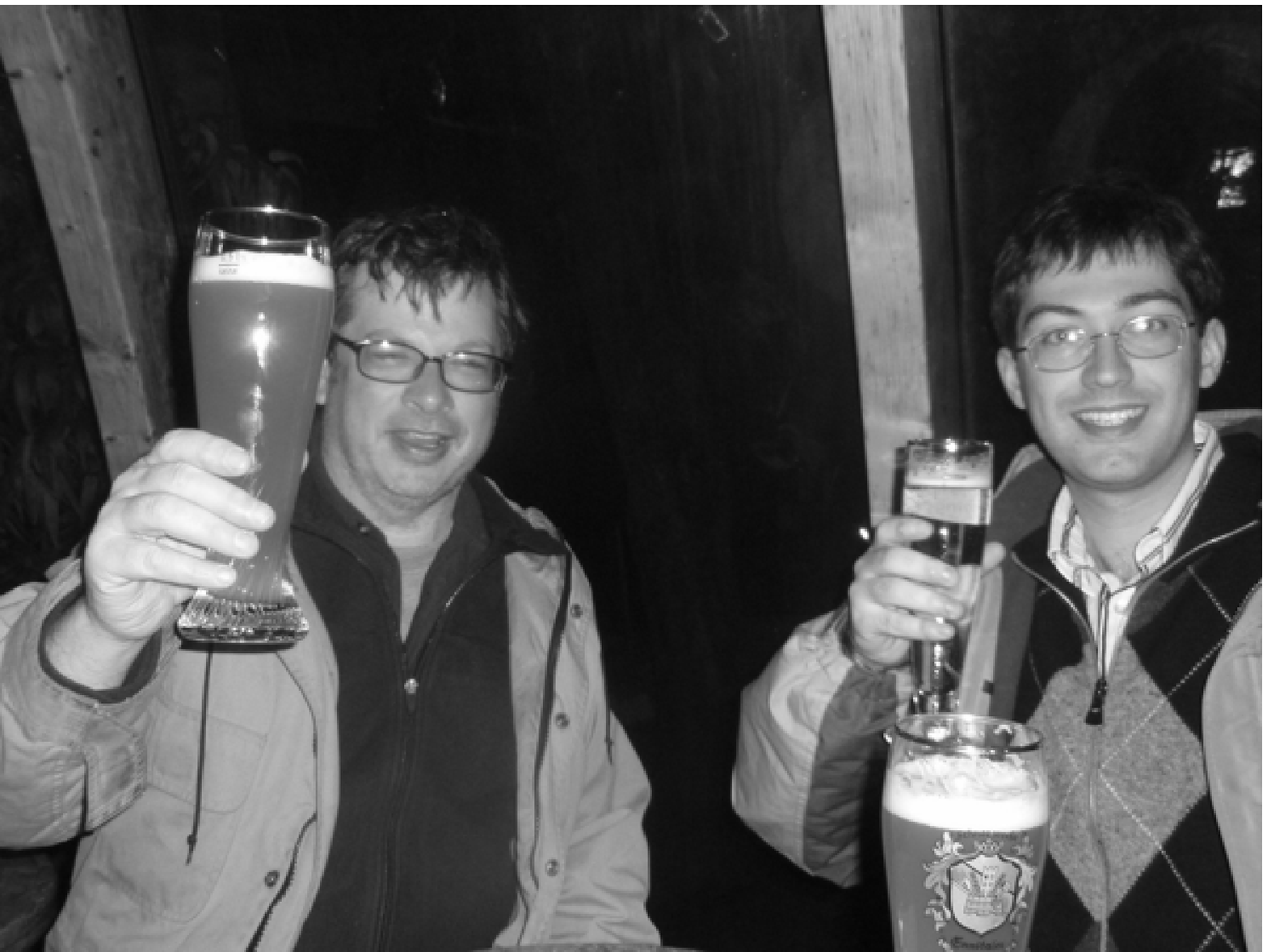}

\caption{Empirical investigations of the number of glasses of good Austrian beer needed in order to convince oneself that a chiral multiplet is present in the hadronic data}\label{beer}
\end{figure*}

One problem which  is obvious at the outset is that the idea is qualitative.  Exactly how close do two resonances need to be before declaring them to be part of a doublet?  Another way to ask the question---which is perhaps appropriate for a winter school in the Styrian Alps---is, ``How many glasses of good Austrian beer does one need to consume before one is convinced that there really is a doublet?''  (See Fig.~\ref{beer}.)

There is also the ``missing state problem''.  Not all resonances that exist have been extracted from the data.  Thus, the fact that not all of the high-lying states have observed doublet partners is not fatal.  The issue is how can we tell if a state needed to fill out a doublet really does not exist or merely has not yet been seen.  In this context, one can fall back on the old saw that ``The absence of proof is not proof of absence.''  However, it is also important to recall the same quote was used by Donald Rumsfeld about Iraq's supposed weapons of mass destruction.  The point here is simply that we do not know about these states.  This issue will become critical when discussing the issue of whether stringy dynamics rather than effective chiral restoration might be the explanation for parity doublets.

There is a third problem with this data: the problem of ``accidental matches''.  The trouble here is that the baryon spectrum is reasonably dense.  One cannot require the masses to be identical; firstly, since the extraction of masses is imprecise, and secondly, because the idea is intrinsically approximate.  This raises the question of whether the near degeneracies are just accidents given the many states in the neighborhood.

\subsection{A statistical analysis\label{statan}}

The issues alluded to above make it very difficult to conclusively show whether or not chiral doublets are a universal feature for highly excited baryons.  In the face of this problem, the MIT group\cite{MITrev} did something rather clever---they lowered the bar.  Rather than asking the hard question of whether parity doublets are ubiquitous, they asked a weaker question; namely, whether there are correlations in mass  between positive parity resonances and negative parity resonances with all other quantum numbers in common.  The virtue of this approach is that it is a necessary condition for the existence of parity doublets while at the same time one can do a systematic statistical study of it.

In doing a statistical study the first step is to decide the appropriate statistical measure.  In principle, one can do standard statistical correlations using masses from the Particle Data Book.  This has many drawbacks, one of which is the missing state problem--the failure to include states which exist but presently unobserved could greatly alter the correlation.  Moreover, it is not clear how to include widths.  To circumvent these problems, an ingenious strategy was adopted\cite{MITrev}.  First  a measure was constructed which is large if positive and negative levels are strongly correlated and small otherwise.  The measure has the feature that missing states will not alter its value and widths are taken into account.  Now, the actual value of this quantity is meaningless.  The trick is to compare the actual value of the quantity to what one expects if the parities of the states had exactly the same properties but their parities were randomly distributed.  This allows one to ask how likely it is that a value of the measure would happen.

The measure constructed was based on an effective spectral density for each parity.  This includes the phenomenological widths from the Particle Data Group (PDG) as well as the masses and is weighted by the confidence that the state is real:
\begin{eqnarray}
\label{spectral}
\rho_{IJS}^{+}(m,\{C\})&=&\sum_{j}\frac{C_{j}W_{j}}{{2}\pi}\frac{\Gamma_{j}}{(m-m_{j})^{2}+\Gamma_{j}^{2}/4}\nonumber\\
\rho_{IJS}^{-}(m,\{C\})&=&\sum_{j}\frac{W_{j}}{{2}\pi}\frac{(1-C_{j})\Gamma_{j}}{(m-m_{j})^{2}+\Gamma_{j}^{2}/4}
\end{eqnarray}
In this construction, the sum is taken  over all states in the Particle Data Book with given $J$, $I$,
and $S$; $W(j)$ takes the values $W_{j} = 1.0, 0.75, 0.50, 0.25$ for $4^{*}$, $3^{*}$,
$2^{*}$, and $1^{*}$ resonances, respectively. The factor $C_{j}$ distinguishes between positive and negative parity states: it  assumes the value $C_{j} = +1$ for positive
parity, and $C_{j}=0$ for negative parity.  Next, the measure $\Omega$ is introduced:
\begin{eqnarray}
\label{statistic}
&{}&\Omega_{IJS}(\{C\}) \equiv  \int dm_{1}dm_{2} {\rm erfc}\left(\frac{|m_{1}-m_{2}|}{\sigma}\right) \nonumber\\
&{}&\times \rho_{IJS}^{+}(m_{1},\{C\})
\rho_{IJS}^{-}(m_{2},\{C\})  \;  ;
\end{eqnarray}
${\rm erfc}(z) =
\frac{2}{\sqrt{\pi}}\int_{z}^{\infty}dte^{-t^{2}}$ is the
complementary error function.   Note that this definition requires the specification of a parameter $\sigma$.  Roughly, $\sigma$ represents the range over which two masses which are similar are considered to have overlapped and hence contribute.  It should be clear from this construction that $\Omega$ does what is needed: it is large when levels are strongly correlated, it includes information both about the widths and the certainty about the states, and missing states don't substantially alter its value.

\begin{figure*}
\centering
\label{nuchist}
\includegraphics[width=4.5in]{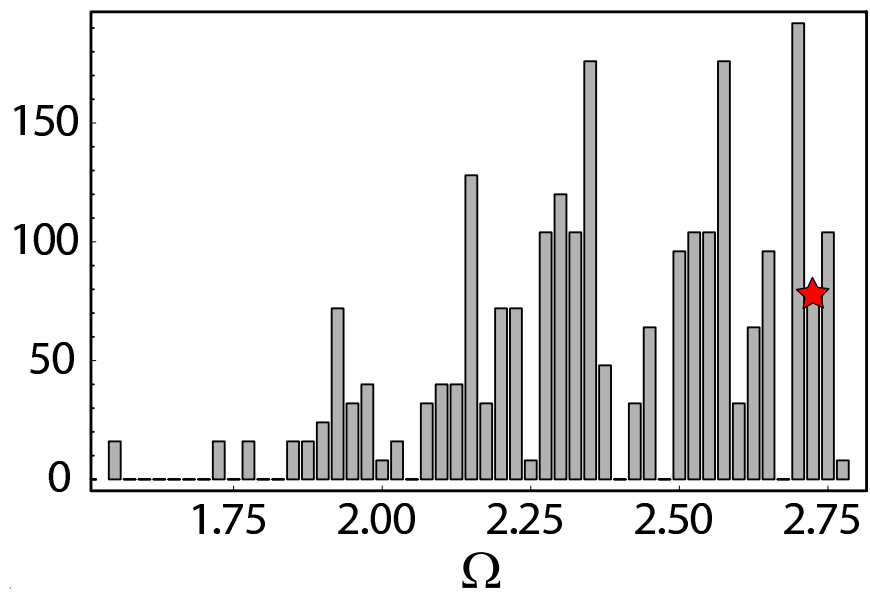}
\caption{A histogram for values of $\Omega$ assuming random parity assignments for the nucleon.  All spins $J=1/2,...,7/2$
are included. The star is the value obtained using real world  parities. $\sigma$ was taken to be 125 MeV in this analysis.  Figure is from ref.~\cite{MITrev} }.
\end{figure*}

As noted earlier, the actual value of $\Omega$ is of no  particular interest.  What is interesting is the question of how likely it is that a particular value would occur if the resonance were as they are in terms of mass and widths but had randomly assigned parities.    Histograms  for nucleons are given in Fig.~\ref{nuchist}.  It is clear that the value is much larger than one would typically expect in the absence of correlations.  Indeed random assignments of parity give a lower value of $\Omega$ 95\% of the time.  Thus, it is reasonable to conclude that there are large correlations.  A similar histogram is shown in Fig.~\ref{nuchist} for the $\Delta$ resonance.  Again the value of $\Omega$ is much larger than would be expected from random assignments of parity; random assignments would give a lower value 86\% of the time.

\begin{figure*}
\centering
\label{deltahist}
\includegraphics[width=4.5in]{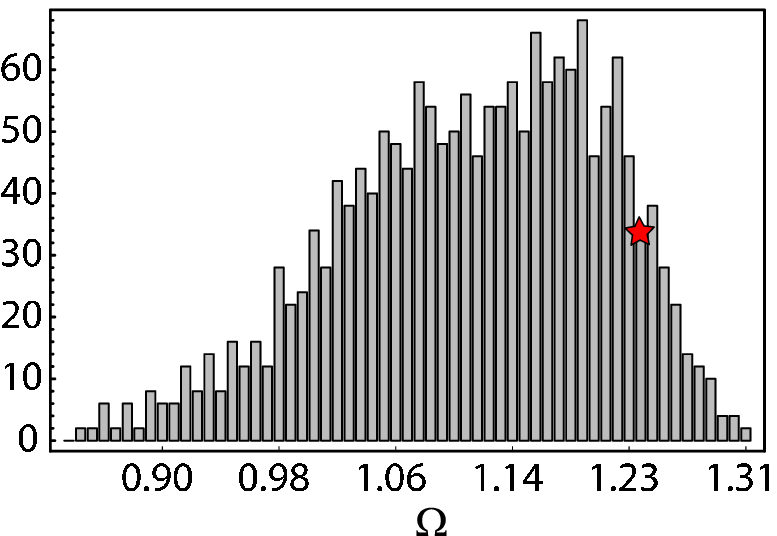}
\caption{A histogram for values of $\Omega$ assuming random parity assignments for the nucleon.  All spins $J=1/2,...,7/2$
are included. The star is the value obtained using real world  parities. Figure from ref.~\cite{MITrev}}.
\end{figure*}

While this statistical analysis does not necessarily prove that high-lying states are always in parity doublets, it does give strong evidence for correlations.

\subsection{Mesons \label{mes}}

If it is true that high-lying baryons fall into parity doublets, an obvious question arises: Is this a feature of baryons, or is it a more general property of hadrons?  When the issue of effective restoration was first being addressed, the data on high-lying mesons  was somewhat sparse and it was hard to reach conclusions.  However, during the early years of this decade a major analysis of the proton-antiproton data from LEAR identified a large number of high-lying mesons\cite{BUGG1,BUGG2,BUGG3,BUGG4,BUGG5,BUGG6}.   It is probably worth noting that these newly discovered states are not listed in the Particle Data Book and accordingly might be considered controversial.  We leave to the reader the question of whether this casts doubt on the states. However, given this situation one should bear in mind the adage ``caveat emptor''.

Once one includes these new high-lying resonances, one finds the same general pattern as with baryons.  Except at very low excitations, mesons of one parity {\it tend} to have a meson of the opposite parity of similar energy.  The quality of the data seems similar to that of the baryons.  As with the baryons, there is the question of how compelling this data is in establishing that the high-lying states do indeed appear in parity doublets.  I would say that the data is at least suggestive.  Perhaps with enough good Austrian beer it is even plausible; see Fig.~\ref{beer}.

\subsection{Chiral symmetry\label{CS}}

One striking feature of QCD is that the masses of the up and down quarks are very light, $\sim 5$ MeV.  This is much smaller than the characteristic masses in hadronic physics.  This suggests that it is useful to think about a world in which the quark masses are strictly zero.  While it is obvious that our experimental friends cannot do measurements in such a world, as a theorist it is extremely attractive to work in the zero quark mass limit---or chiral limit as it is called.  The idea is to compute in this relatively tractable limit and then include quark mass effects as perturbative corrections.

In this zero quark mass world, QCD is invariant under the following transformations among the quarks:
\begin{eqnarray}
q & \rightarrow & e^{i \vec{\theta}_v \cdot \vec{\tau}} q \; \; \; \; {\rm vector}   \nonumber \\
q & \rightarrow & e^{i \gamma_5 \vec{\theta}_A \cdot \vec{\tau}} q \; \; \; \;  {\rm axial} \; ,
\end{eqnarray}
where $\vec{\tau}$ are Pauli matrices.
An alternative way to parameterize these transformations is
\begin{eqnarray}
q & \rightarrow & e^{i (1-\gamma_5) \vec{\theta}_L \cdot \vec{\tau}} q \; \; \; \; {\rm left} \nonumber \\
q & \rightarrow & e^{i (1+\gamma_5) \vec{\theta}_R \cdot \vec{\tau}} q \; \; \; \;  {\rm right} \; .
\end{eqnarray}
In this parameterization the transformations are for the left-handed and right-handed quarks separately---hence the name ``chiral''.  In the chiral representation, it is clear that QCD in the massless limit is invariant under an $SU(2) \times SU(3)$ symmetry.  The only term in the QCD lagrangian {\it not} invariant under these transformations are the {\it very small} mass terms.

\begin{figure*}
\centering
\includegraphics[width=6in]{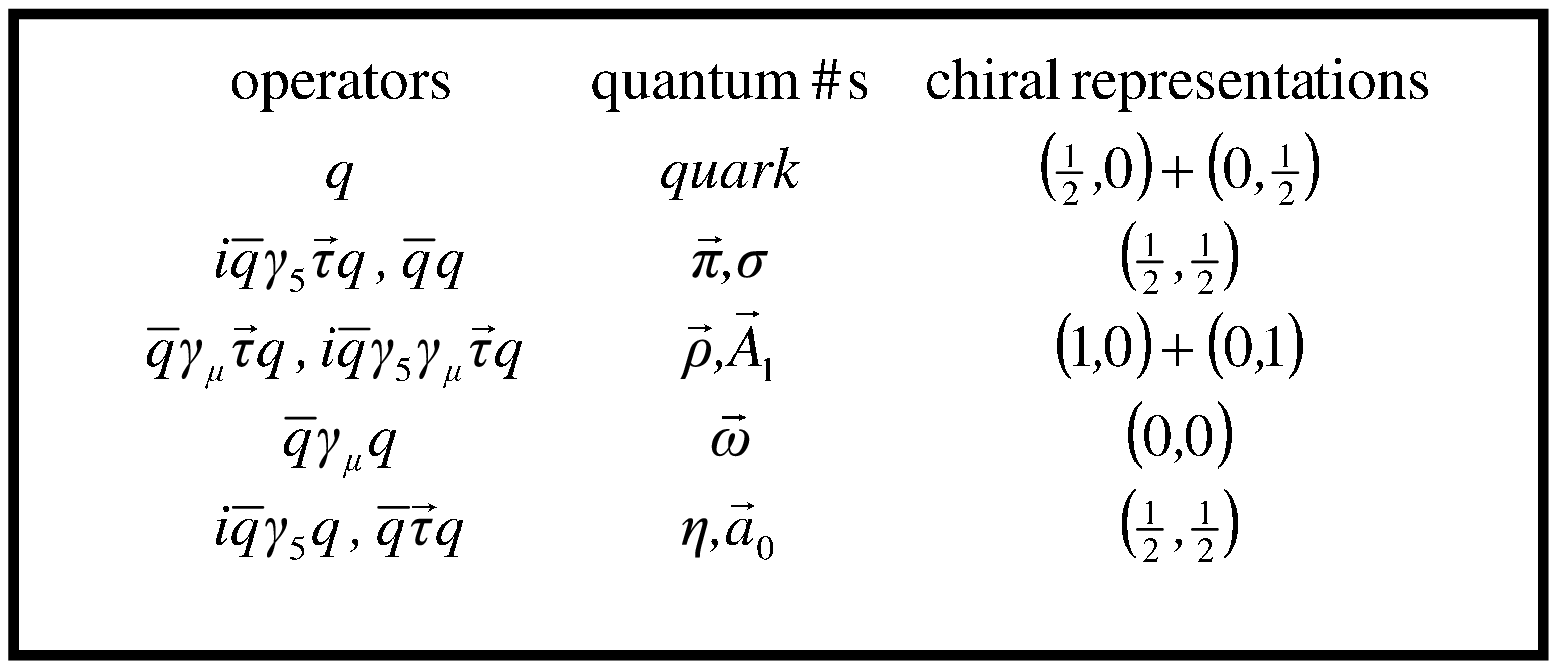}
\caption{Some QCD operators and their chiral representations.  Note typically two representations of chirality are combined into a single chiral/parity representation.}
\label{operators}
\end{figure*}

These chiral transformations form a group.  Representations of the group can be given in terms of the representations of $SU(2)_L$ and $SU(2)_R$. For example, $(\frac{1}{2},0)$  means the left-handed quarks transform as an isodoublet while the right-handed quarks transform as an isosinglet.  Typical operators in QCD transform into each other under chiral rotations.  Thus the operators fall into chiral representations.  However, since the operators have well-defined parity, they actually correspond to {\it two} chiral representations.  We label these as chiral/parity representations.  Some examples are given in Fig.~\ref{operators}.

\

 The fact that any given chiral multiplet---except the trivial one---breaks parity is essential to what follows.

\subsection{Spontaneous chiral symmetry breaking}

\begin{figure}
\centering
\includegraphics[width=3.in]{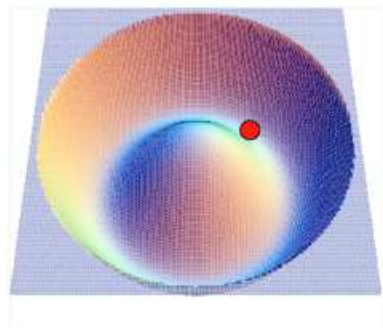}
\caption{A Mexican hat potential.  The dot represents {\it one} of the ground states. }
\label{MexHat}
\end{figure}

Although chiral symmetry is, to good approximation, a symmetry of the QCD Lagrangian, it is {\it not} a good symmetry of the QCD ground state---{\it i.e.} its vacuum.  This situation is called spontaneous symmetry breaking.  A classical way to visualize this is in terms of a particle in a ``Mexican hat'' potential as in Fig.~\ref{MexHat}.

Of course, in quantum mechanics for a finite system this situation cannot occur---the ground state will be a wave function which symmetrically samples all of these.  However, for infinite systems this is not true and spontaneous symmetry breaking is possible.  The 2008 Nobel Prize in physics went to Nambu\cite{Nobel} precisely for pointing out that in field theory spontaneous symmetry breaking can occur.  The symmetry he discussed in this seminal work was chiral symmetry in strong interactions.

Now there is an obvious question---How do we know that QCD breaks chiral symmetry?  The standard argument is the existence of Goldstone modes---the pions.  Goldstone showed long ago that for every spontaneously broken continuous symmetry generator, there will be a massless particle.  (It can, of course, be eaten in a Higgs mechanism.)  The pions in real QCD are pseudo-Goldstone bosons.  ``Pseudo'' because they are not massless but only {\it nearly} massless compared to other masses in strong interactions.  This happens because chiral symmetry is not exact---there is explicit chiral symmetry breaking due to the quark masses.  Moreover, there is additional evidence besides the mass that pions are  pseudo-Goldstone bosons.  One can show that in the chiral limit and at zero external momenta, Goldstone bosons decouple; their scattering amplitudes go to zero---even after phase space effects have been divided out.  While not zero, pion couplings at low momentum are empirically very small---much smaller than one would estimate via dimensional analysis using the scales in strong interaction physics.   This is precisely what one would expect if pions were pseudo-Goldstone bosons.

There is another way to see that chiral symmetry must be spontaneously broken.  If the vacuum state had been invariant under chiral symmetry, then hadronic states would have to form chiral multiplets---or more precisely, chiral/parity multiplets of degenerate states.  Since all multiplets which are not pure isosinglets contain states of both parities, an unbroken chiral symmetry would requite that, for example, the nucleon would have a nearly degenerate state (exactly degenerate in the massless quark limit) with the opposite parity.  However, this is not, in fact, the case.  The lightest negative parity nucleon with spin 1/2 is the $N^*(1535)$ which is nearly 600 MeV heavier than the nucleon.

The original motivation about chiral restoration is based precisely on this last fact.  One can invert the argument.  If the absence of nearly degenerate states of opposite parity is evidence for the breaking of spontaneous chiral invariance, it is not unreasonable to ask whether the {\it presence} of opposite parity degenerate states is a signature of some kind of effective chiral restoration.  Glozman's conjecture was that this is what is indeed happening.

\section{The Nature of the Conjecture}

Before going forward it is important to clarify precisely what the conjecture {\it is}; {\it i.e.} what precisely is meant by effective chiral restoration for high-lying hadrons.  There has been significant confusion about this in the literature.  In part this may be due to the fact that the discussions in the earliest papers\cite{G1} used language in a somewhat unfortunate way.  The confusion about the meaning of the conjecture has vexed the field for a considerable period.  The key thing to realize is that the conjecture does {\it not} imply that the high-lying states are somehow in a different phase than the low-lying states.  Rather, the conjecture is based on the notion that there is some mechanism or mechanisms which are responsible for spontaneous chiral symmetry breaking.  Effective chiral restoration simply means that as one goes to higher masses, the states become increasingly insensitive to the dynamics responsible for chiral symmetry breaking\cite{CG1,GlozmanReview}.  Thus, although the system remains in a chirally broken phase, the {\it effect} of  the spontaneous symmetry breaking on these states becomes small.

An obvious requirement for such a scheme to explain parity doublets in the data is for the hadrons still to be narrow enough to be identifiable as hadrons when the masses get narrow enough for effective restoration occurs.

\subsection{chiral multiplets versus parity doublets for baryons \label{chiplet}}

There is an important theoretical issue associated with effective chiral restoration.  While as noted above, effective restoration implies states with the same quantum numbers and opposite parity should be nearly degenerate, it actually implies more: it implies that high lying hadrons for chiral multiplets of degenerate states\cite{Duck}.  In general, these are larger than parity doublets.

Consider, for example, the case of baryons\cite{CJ,CG1}.  Suppose that one were to make the naive assumption that all baryons have the quantum numbers of three chiral quarks.  Then the possible chiral/parity multiplets are given by
\begin{eqnarray}
(1/2,0)& + &(0,1/2)\\
(1/2,1)& + &(1,1/2)\\
(3/2,0)& + &(0,3/2) \; .
\end{eqnarray}
The first of these is a parity doublet of nucleons, the second a multiplet of positive and negative parity nucleons and $\Delta$s, while the third is a parity doublet of $\Delta$s.  The second class of multiplets is something larger than just parity doublets.

 We can ask whether the data falls into multiplets like this as would be expected if effective restoration is the explanation for the phenomenon or whether there is only parity doubling.  Before doing this, it is useful to note why the preceding analysis of representations is naive.  The notion that a baryon is made of three quarks is from the quark model rather than QCD.  In principal, nothing in QCD prevents baryons from having the quantum numbers of, say, four quarks and one anti-quark or five quarks and two anti-quarks.  If one allows such states the number of possible representations  increases.  On the other hand, so far as we know the non-chiral quantum numbers of the observed hadrons are consistent with them being ``non-exotic''; {\it i.e.}, having the same quantum numbers as a naive quark model (three quarks for a baryon, quark-antiquark for the mesons).  Thus in making a preliminary analysis it probably makes sense to see if these naive representations are sufficient to explain the data.  If not, one can consider larger representations not obtainable from three quarks.  As it happens, there is no need to do this.

If one looks at the data, one can tentatively assign the high-lying baryons into multiplets\cite{GlozmanReview} of either $(1/2,0) +(0,1/2$ or $(1/2,1) +(1,1/2)$.  First consider the putative $(1/2,0) +(0,1/2)$ states:
\begin{eqnarray*}
 J & = &\frac{1}{2}: N^+(1710), N^-(1650)\\
 J& = &\frac{3}{2}: N^+(1720), N^-(1700)\\
  J& = &\frac{5}{2}: N^+(1680), N^-(1675)\\
J& = &\frac{9}{2}: N^+(2220), N^-(2250)
\end{eqnarray*}

The putative $(1/2,1) +(1,1/2)$  baryons are above 1.9 GeV.:
\begin{eqnarray*}
J & = &\frac{1}{2} :\\
 &{}& N^+(2100)^{*},N^-(2090)^{*}\\ &{}&\Delta^+(1910),\Delta^-(1900)^{**}\\ \\
 J &=& \frac{3}{2} :\ \\
&{}& N^+(1900) {**},~N^-(2080)^{**} \\ &{}& \Delta^+(1920),\Delta^-(1940)^{*}\\
 J &=&\frac{5}{2} :\\\
&{}& N^+(2000)^{**},N^-(2200)^{**} \\ &{}& \Delta^+(1905),\Delta^-(1930)\\  \\
 J &= &\frac{7}{2}: \\
&{}&  N^+(1990)^{**}, N^-(2190)\\ &{}&  \Delta^+(1950)  \Delta^-(2200)^{*}\\  \\
  J &=&\frac{9}{2}:\ \\
 &{}& ~N^+(2220),N^-(2250)\\ &{}& \Delta^+(2300)^{**} \Delta^-(2400)^{**}\\  \\
  J & =& \frac{11}{2} :\\
&{}& ?,N^-(2600) \\ &{}&  \Delta^+(2420),?;\\ \\
  J &=&\frac{13}{2} :  \\
 &{}& N^+(2700)^{**}, ?,? \\&{}&  \Delta^-(2750)^{**}\\  \\
  J&=&\frac{15}{2} : \\
 &{}& ?,  ? \\ &{}& \Delta^+(2950)^{**},?.
\end{eqnarray*}
The stars are according to the Particle Data Group assignment and correspond to questionable states.  The question marks correspond to baryons which have not been observed but would be needed to fill out the multiplets.

How compelling is this data that the baryons fall into chiral/parity multiplets? Clearly the data cannot be considered compelling until the states with one or two stars are verified and the missing states identified in the scattering data.  Moreover, the splitting within the multiplets do not follow an obvious pattern.  Nevertheless, the data does seem rather suggestive.  Perhaps with enough good Austrian beer, it may even seem rather plausible.  See Fig.~\ref{beer}.

One can, of course, do a similar analysis for the mesons.  In doing this it is important to recall\cite{CJ} that there is no unique way to assign chiral transformation properties given usual quantum numbers.  Thus a QCD operator which creates a  $\rho$ can transform as $\overline{q} \vec{\tau} \gamma_{\mu} q$ and be part of a $(0,1)+(1,0)$ multiplet in a chirally restored phase.  But it could also be $\overline{q} \vec{\tau} (\partial_\mu q)$ and be part of a $(1/2,1/2)$ multiplet.  Putative chiral multiplet for high-lying mesons have been identified\cite{G3,GlozmanReview}.  In doing this one needs to identify {\it which} chiral multiplet one assigns a particular meson.  The states studied include both states in the PDG and recently identified states seen in proton-antiproton collisions. Let us start by looking at $J=0$ meson which can be taken to be in the $(1/2,1/2)$ chiral representation:
\begin{center} {\bf (1/2,1/2)}\\

{$\pi(0,1^{--}) ~~~~~~~~~~~~~~~~~~~~~~~ f_0(0,0^{++})$}\\
\medskip
{$11300~\pm~100 ~~~~~~~~~~~~~~~~~~~~~~~~ 1370~\pm~{}^{130}_{170}$}\\
{~$1812~\pm~14 ~~~~~~~~~~~~~~~~~~~~~~~~~ 1770~\pm~12$}\\
{~$2070~\pm~35 ~~~~~~~~~~~~~~~~~~~~~~~~~ 2040~\pm~38$}\\
{~$2360~\pm~25 ~~~~~~~~~~~~~~~~~~~~~~~~ 23370~\pm~14$}\\
\end{center}
Next consider the states with $J>0$. First let us look at $J=2$ states which appear to have the most complete data:
\begin{center}  {\boldmath $(1/2,1/2)_a$}\\

{$\pi_2(1,2^{-+})~~~~~~~~~~~~~~~~~~~~~~~f_2(0,2^{++})$}\\
\medskip
{$2005 \pm 15   ~~~~~~~~~~~~~~~~~~~~~~~~2001 \pm 10$}\\
{$2245 \pm 60   ~~~~~~~~~~~~~~~~~~~~~~~~2293 \pm 13$}\\

\bigskip
{\boldmath  $(1/2,1/2)_b$}\\

{$a_2(1,2^{++})~~~~~~~~~~~~~~~~~~~~~~~ \eta_2(0,2^{-+})$}\\
\medskip
{ $2030 \pm 20  ~~~~~~~~~~~~~~~~~~~~~~~~2030 ~\pm ~?$}\\
{ $2255 \pm 20 ~~~~~~~~~~~~~~~~~~~~~~~~2267 \pm 14$}\\

\bigskip
{\bf (0,1)+(1,0)}\\

{$a_2(1,2^{++})~~~~~~~~~~~~~~~~~~~~~~~\rho_2(1,2^{--})$}\\
\medskip
{ $1950^{+30}_{-70}~~~~~~~~~~~~~~~~~~~~~~~~~~~1940 \pm 40$}\\
{ $2175 \pm 40  ~~~~~~~~~~~~~~~~~~~~~~~~2225 \pm 35$}\\
\end{center}

The data sets for the $J=1$ and $J=3$ mesons have ``missing states'' from the point of view of chiral restoration.  As noted earlier it is not clear whether this means the states do not exist or merely that they have not been observed:  ``The absence of proof is not proof of absence''.
First the data for $J=1$:
\begin{center}
{\bf (1/2,1/2)}\\

{$\omega (0,1^{--})~~~~~~~~~~~~~~~~~~~~~~~b_1(1,1^{+-})$}\\
\medskip
{$1960 \pm 25   ~~~~~~~~~~~~~~~~~~~~~~~~1960 \pm 35$}\\
{$2205 \pm 30   ~~~~~~~~~~~~~~~~~~~~~~~~2240 \pm 35$}\\

\bigskip
{\bf (1/2,1/2)}\\

{$h_1(0,1^{+-})~~~~~~~~~~~~~~~~~~~~~~~\rho(1,1^{--})$}\\
\medskip
{$1965 \pm 45   ~~~~~~~~~~~~~~~~~~~~~~~~1970 \pm 30$}\\
{$2215 \pm 40   ~~~~~~~~~~~~~~~~~~~~~~~~2150 \pm ~?$}\\

\bigskip
{\bf (0,1)+(1,0)}\\

{$a_1(1,1^{++})~~~~~~~~~~~~~~~~~~~~~~~\rho (1,1^{--})$}\\
\medskip
{$1930^{+30}_{-70} ~~~~~~~~~~~~~~~~~~~~~~~~~1900 ~\pm ~?$}\\
{$ 2270^{+55}_{-40} ~~~~~~~~~~~~~~~~~~~~~~~~~2265 \pm 40$}\\

\end{center}

Finally, the multiplets for $J=3$.
\begin{center}{\bf (0,0)}\\

{$\omega_3(0,3^{--})~~~~~~~~~~~~~~~~~~~~~~~f_3(0,3^{++})$}\\
\medskip
{$~~~~~?~~~~~   ~~~~~~~~~~~~~~~~~~~~~~~~~2048 \pm 8$}\\
{$2285 \pm 60   ~~~~~~~~~~~~~~~~~~~~~~~~2303 \pm 15$}\\

\bigskip
{\bf (1/2,1/2)}\\

{$\omega_3 (0,3^{--})~~~~~~~~~~~~~~~~~~~~~~~b_3(1,3^{+-})$}\\
\medskip
{$1945 \pm 20   ~~~~~~~~~~~~~~~~~~~~~~~~2032 \pm 12$}\\
{$2255 \pm 15   ~~~~~~~~~~~~~~~~~~~~~~~~2245 \pm ~?$}\\

\bigskip
{\bf (1/2,1/2)}\\

{$h_3(0,3^{+-})~~~~~~~~~~~~~~~~~~~~~~~\rho_3(1,3^{--})$}\\
\medskip
{$2025 \pm 20   ~~~~~~~~~~~~~~~~~~~~~~~~1982 \pm 14$}\\
{$2275 \pm 25   ~~~~~~~~~~~~~~~~~~~~~~~~2260 \pm 20$}\\

\bigskip
{\bf (0,1)+(1,0)}\\

{$a_3(1,3^{++})~~~~~~~~~~~~~~~~~~~~~~~\rho_3 (1,3^{--})$}\\
\medskip
{$2031 \pm 12  ~~~~~~~~~~~~~~~~~~~~~~~~ 2013 \pm 30$}\\
{$2275 \pm 35 ~~~~~~~~~~~~~~~~~~~~~~~~~2300^{~+~50}_{~-~80}$}\\

\end{center}

There is insufficient data for higher spin mesons to draw conclusions.  Now, does the preceding data provide compelling evidence for effective restoration?  Again my judgement is that the data is suggestive, rather than compelling.  Again, with enough good Austrian beer, it may even seem quite plausible.  See Fig.~\ref{beer}.

\subsection{The axial anomaly}

\begin{figure}
\centering
\includegraphics[width=3.in]{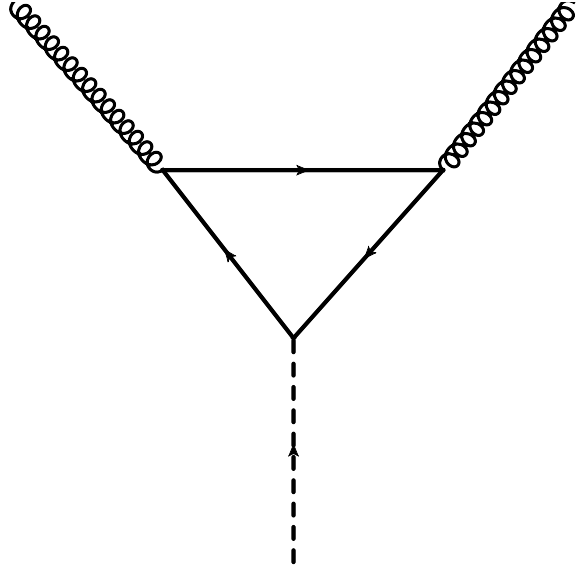}
\caption{The triangle graph responsible for the axial anomaly.  The dashed line represents an axial current. }
\label{anom}
\end{figure}

Before proceeding it is useful to review some facts about the axial anomaly. Note that at the level of the Lagrangian, QCD has more symmetry than just the $SU(2)_L \times SU(2)_R$ chiral symmetry.  There is also an axial $U(1)$ symmetry.  The $U(1)$ axial transformation is simply
\begin{equation}
q \rightarrow e^{i \theta} q \; .
\end{equation}
In the absence of quark masses, the Lagrangian is invariant under this transformation, implying a conserved axial current.  However, at the quantum level this symmetry is broken due to an anomaly---the act of quantizing theory requires some method of regularizing divergences.  It turns out that to keep the theory gauge invariant, the only way to regulate breaks the symmetry\cite{ANOMALY1,ANOMALY2}.  Fortunately it breaks it in a known way. In the massless quark limit this is given by:
\begin{equation}
\partial_{\mu}\vec{J}^{\,\mu}_{5} =  -i\frac{2g^{2} N_f}{16\pi^{2}}{\rm Tr}\tilde
F^{\mu\nu}F_{\mu\nu}
\end{equation}
where $\vec{J}^{\,\mu}_{5}$ is the $U(1)$ axial current.  This can be seen via perturbative computation of the triangle graph; see Fig.~\ref{anom}.  It can be shown that there are no corrections at higher order.
The anomaly acts like an explicit symmetry breaking term.  Thus even though the ``would be'' $U(1)$ axial symmetry is spontaneously broken in the $\langle \overline{q}q \rangle \ne 0$, there is no Goldstone mode associated with it.

\subsection{The axial anomaly and parity doubling}

Like the $SU(2)_L \times SU(2)_R$ chiral symmetry, the $U(1)$ axial symmetry mixes states of even and odd parity.  Thus one might suspect that it could play a role in the phenomenon of parity doubling.  Indeed, the MIT group\cite{Duck,jaffe,MITrev} suggested that ``suppression of $U(1)_A$ violation'' may be responsible for parity doubling.

It is not immediately clear what ``suppression of $U(1)_A$ violation'' means.  On the one hand, it sounds like it means that the effect of the anomaly becomes unimportant at high masses leading to parity doubling.  If that is what the idea means then it is easy to see that it is wrong.  The key point is that the $U(1)_A$ symmetry is not only anomalously broken, it is also spontaneously broken.  That is, the chiral condensate  $\langle \overline{q}q \rangle$ breaks $U(1)_A$ symmetry as well as the usual chiral symmetry.  Since we know from the usual analysis of chiral symmetry that $\langle \overline{q}q  \rangle \ne 0$ (even in the massless quark limit), we know that the $U(1)_A$ symmetry is also spontaneously broken.  Since spontaneous symmetry breaking will also destroy parity doublets, parity doublets will not occur---even if $U(1)_A$ violation vanishes.

This implies that {\it even} if  anomalous $U(1)_A$ violation is suppressed for high-lying states, parity doublets will still require that the effects of spontaneous breaking are also suppressed.  Since these effects also lead to the breaking of chiral multiplets it appears as though  parity doublets due to the suppression of the anomalous $U(1)_A$ violation only occur if there are {\it also}  chiral multiplets.

\begin{figure}
\centering
\includegraphics[width=2.5in]{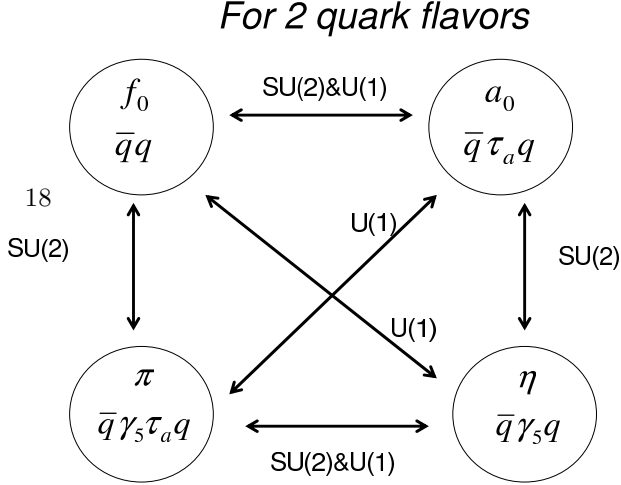}
\caption{If all symmetries are unbroken these are all  degenerate; if SU(2) is spontaneously broken and no anomaly contribution: $\pi$, $\eta$  are degenerate Goldstone bosons while $f_0$ , $a_0$ are both massive.}
\label{anom-mes}
\end{figure}

This is easily illustrated in the meson sector: consider scalar and pseudoscalar mesons.  These are connected by $SU(2)$ and $U(1)$ axial symmetry as in Fig.~\ref{anom-mes}.  Note that the spontaneous symmetry breaking of the usual axial currents split the $\pi$ from the $a_0$ its would be parity doublet as well as from its chiral partner the $f_0$.

It is equally easy to see that even if the anomaly plays no role, when chiral symmetry is spontaneously broken, parity doublets do not automatically occur.  The simplest way to see this is in the context of the Skyrme model\cite{Skyrme1,Skyrme2,Skyrme3}. It is easy to generalize the model\cite{SkyrmeU1} to include $U(1)_A$ degrees of freedom by basing it on a $U(2)$ matrix rather than an $SU(2)$ matrix and thereby including an $\eta$ (perhaps better called an $\eta'$) degree of freedom. In its simplest incarnation the model incorporates chiral symmetry and its spontaneous breaking but does {\it not} include anomalous $U(1)_A$ breaking (which can be added in if one likes as a $1/N_c$ correction). The model does {\it not}  have parity doubling.  Thus, the absence of anomalous  $U(1)_A$ violation by itself does not yield parity doubling. The reason is clear---at least in the Skyrme model.  Now, of course one can argue that the Skyrme model is not QCD so  one must use caution when drawing conclusions.  Nevertheless, the model does properly encode the symmetries of QCD and thus illustrates that  the lack of anomalous  $U(1)_A$ violation is compatible with the absence of parity doubling.  Therefore, merely turning off the effects of the anomaly does {\it not} automatically induce parity doublets.  The reason is as noted before---spontaneous breaking of the $U(1)_A$ spoils parity doubling and the spontaneous breaking of the $SU(2)_L \times SU(2)_R$ automatically implies spontaneous breaking of the $U(1)_A$.

The preceding argument shows that viewed naively a ``suppression of $U(1)_A$ violation'' cannot by itself be the cause of parity doublets.  However, the MIT group's explanation for ``suppression of $U(1)_A$ violation'' is a bit more subtle.  It is based on the equations of motion\cite{MITrev}.  For simplicity let us restrict attention to the massless quark limit:
\begin{eqnarray}
i\dot{Q}_5 &= &[H,Q_5]= \int d^3 x \partial_mu J_5^\mu =D_5 \nonumber \\
D_5 &\equiv& \int d^3 x \frac{2g^{2} N_f}{16\pi^{2}}{\rm Tr}\tilde
F^{\mu\nu}F_{\mu\nu},
\end{eqnarray}
where $D_5$ is the space integral of the divergence of the current.  Now let us take matrix elements of this equation between baryon states of opposite parities:
\begin{equation}
\langle B'| [H,Q_5] |B \rangle = \langle B'| D_5|B \rangle \; \label{com1} .
\end{equation}
Now suppose that ``suppression of $U(1)_A$ violation'' means that for high-lying baryons the right-hand side of Eq.~(\ref{com1}) is small.  For simplicity, let us take it to be zero: $\langle B'| [H,Q_5] |B \rangle =0$.  On the other hand, one can always write
\begin{eqnarray}
&{}&\langle B'| [H,Q_5] |B \rangle = \nonumber \\
&{}&\left(  M(B') -M(B) \right)\langle B'| Q_5 |B \rangle \; . \label{com2}
\end{eqnarray}
Next assume that the right-hand side of Eq.~(\ref{com2}) is not zero. Then the vanishing $\langle B'| [H,Q_5] |B \rangle =0$ implies that $M(B')=M(B)$---in  other words, the suppression of $U(1)_A$ violation implies degeneracy of the baryons, {\it i.e.}, parity doublets.  This {\it appears} to show a deep connection between a dynamical suppression of the effects of the axial anomaly and the existence of parity doublets.

However, appearances can be deceiving.  In the first place this analysis depends on a critical assumption besides ``suppression of $U(1)_A$ violation'' ({\it i.e.} small matrix elements of $D_5$ between high-lying states). It also depends on the righthand side of Eq.~(\ref{com2}) {\it not} being small.  However, this assumption is {\it ad hoc} and not generally satisfied.  For instance, suppose we lived in a world where the anomaly was shut off entirely---one can, for example, think of a large $N_c$ world.  In such a world the right-hand side of Eq.~(\ref{com2}) is identically zero by construction.  But we know from the Skyrme model that parity doublets do not automatically emerge in that world, ergo the right-hand side of Eq.~(\ref{com2}) vanishes. Moreover, if the anomalous $U(1)_A$ violation is generally suppressed for high-lying states and $\langle B'| Q_5 |B \rangle$ is generically not small, then {\it all} high-lying baryons of opposite parity and the same quantum numbers will be degenerate---not just states in the same doublet.  Thus for this scheme to work, $\langle B'| D_5|B \rangle $ must be large for two baryons within a parity doublet and small if the two baryons are in different doublets.  Unless, one has an explanation for why this happens one has explained very little.

More importantly, there is a sense that this analysis explains nothing.  While it appears that fundamental physics associated with the anomaly is at play here, this is not really the case. Consider {\it any} non-conserved vector ``current'', $C_\mu$, with $\partial^\mu C_\mu= \Theta$.  It is easy to see that
\begin{eqnarray}
&{}&\langle B'| [H,Q_C] |B \rangle = \langle B'| \int d^3 x \Theta |B \rangle \nonumber \\ &{}&\left(  M(B') -M(B) \right)\langle B'| Q_C |B \rangle \
\end{eqnarray}
Now it is clear that the existence of a degeneracy such as in parity doubling implies that the matrix element $\langle B'| \int d^3 x \Theta |B \rangle$ must vanish.   The key point here is that vanishing of $\langle B'| \int d^3 x \Theta |B \rangle$ is completely independent of one's choice of a non-conserved current.  While the matrix element of the divergence of the $U(1)_A$ current vanishes---so do the matrix element of {\it every} current with these quantum numbers.  Thus, for example, if one took
\begin{equation}
C_\mu= \overline{q} \gamma_5 (\partial_mu F^2) q \; ,
\end{equation}
its divergence will have a vanishing matrix element between any degenerate state.  It looks as though a ``suppression of the violation of  $C$ conservation'' is happening and that this is responsible for the degeneracy.  However, there is no dynamics in this statement. It does not {\it explain} why the states are degenerate---rather, it is just a restatement of the fact there is a degeneracy.  The  ``suppression of $U(1)_A$ violation'' is no different.

\subsection{Effective restoration of chiral and $U(1)_A$ symmetries}
In the previous section, it was shown that the suppression of the anomalous $U_A$ violations in high-lying matrix elements does not explain the parity doublets in the absence of effective chiral restoration.  However, it is possible that effective chiral restoration occurs at the same time  that the anomalous $U(1)_A$ axial symmetry is also effectively restored (in the sense that matrix elements of its divergence are small). How would this affect the pattern of degeneracies?

For the baryons, it turns out that has no effect on the degeneracy patterns: all baryon states connected by $U(1)_A$ axial transformations are also connected by ordinary chiral restoration.  Thus, for example, if baryons are in the $(1/2,0) + (0,1/2)$ chiral/parity multiplet, the states are positive and negative parity nucleon states.  Now the $U(1)_A$ transformation simply takes the positive and negative parity nucleons into each other---no new states are coupled in.

The situation with mesons is quite different.  Looking at Fig.~\ref{anom-mes} we see that if one has both effectively unbroken chiral and $U(1)_A$ symmetries then the pion, the eta meson, the $f_0$ and $a_0$ would all be degenerate.  Thus instead of having two distinct chiral $(1/2,1/2)$ multiplets ($\pi$, $f_0$ and $\eta$, $a_0$) we have one larger multiplet.  A similar situation occurs for meson with other spins.

It is an empirical question as to whether the pattern of degeneracies  support the hypothesis that both chiral and $U(1)_A$ are effectively restored for high mass states.  Glozman studied the data and concluded there was evidence for this\cite{G2,GlozmanReview}.   This was done by asking which additional states would be degenerate due to $U(1)_A$ given the  the same identifications of chiral multiplets as in Sec. \ref{chiplet}.

\begin{center}{\bf J=1}\\

{$\omega(0,1^{--})~~~~~~~~~~~~~~~~~~~~~~~h_1(0,1^{+-})$}\\
\medskip
{$1960 \pm 25   ~~~~~~~~~~~~~~~~~~~~~~~~1965 \pm 45$}\\
{$2205 \pm 30   ~~~~~~~~~~~~~~~~~~~~~~~~2215 \pm 40$}\\

\bigskip

{$b_1 (1,1^{+-})~~~~~~~~~~~~~~~~~~~~~~~\rho(1,1^{--})$}\\
\medskip
{$1960 \pm 35   ~~~~~~~~~~~~~~~~~~~~~~~~1970 \pm 30$}\\
{$2240 \pm 35   ~~~~~~~~~~~~~~~~~~~~~~~~2150 \pm ~?$}\\

\bigskip
{\bf J=2}\\

{$f_2(0,2^{++})~~~~~~~~~~~~~~~~~~~~~~~\eta_2(0,2^{-+})$}\\
\medskip
{$2001 \pm 10  ~~~~~~~~~~~~~~~~~~~~~~~~2030 ~\pm ~?$}\\
{$2293 \pm 13   ~~~~~~~~~~~~~~~~~~~~~~~~2267 \pm 14$}\\

\bigskip

{$\pi_2(1,2^{-+})~~~~~~~~~~~~~~~~~~~~~~~a_2 (1,2^{++})$}\\
\medskip
{$2005 \pm 15  ~~~~~~~~~~~~~~~~~~~~~~~~ 2030 \pm 20$}\\
{$2245 \pm 60 ~~~~~~~~~~~~~~~~~~~~~~~~ 2255 \pm 20 $}\\

\bigskip

{\bf J=3}\\

{$\omega_3(0,3^{--})~~~~~~~~~~~~~~~~~~~~~~~h_3(0,3^{+-})$}\\
\medskip
{$1945 \pm 20   ~~~~~~~~~~~~~~~~~~~~~~~~2025 \pm 20$}\\
{$2255 \pm 15   ~~~~~~~~~~~~~~~~~~~~~~~~2275 \pm 25$}\\

\bigskip

{$b_3 (1,3^{+-})~~~~~~~~~~~~~~~~~~~~~~~\rho_3(1,3^{--})$}\\
\medskip
{$2032 \pm 12   ~~~~~~~~~~~~~~~~~~~~~~~~1982 \pm 14$}\\
{$2245 ~\pm ~?   ~~~~~~~~~~~~~~~~~~~~~~~~2260 \pm 20$}\\

\end{center}

Again, one can ask whether the data is compelling.  Again, I would say the data is suggestive rather than compelling.  Again, with enough good Austrian beer, it may seem plausible.  See Fig.~\ref{beer}.

Thus the conjecture can be stated as follows: high-lying hadron states exhibit approximate chiral and $U(1)$ axial restoration in the sense that the states are insensitive to the dynamics responsible for spontaneous chiral symmetry breaking and the axial anomaly.  This leads to a pattern of nearly degenerate states including states of opposite parities.

\section{Theory Issues}

The idea of effective chiral restoration typically elicits one of two reactions on first introduction to a theorist:  Either it appears to be trivially correct or obviously wrong.

The first reaction stems from the following reasoning.  Clearly there is a natural scale associated with chiral symmetry breaking.  To the extent that one studies states which are much heavier than this scale, then the states will not feel the effects associated with chiral symmetry breaking and hence one will see effective chiral multiplets.  This argument is very plausible on the surface, but for reasons which will become clear later, the issue is somewhat subtle.  While the conjecture may ultimately turn out to be correct, the simple argument based on scales is not sufficient to show it.

The second reaction is the idea is crazy and must be wrong.  I think this second reaction stems at least in part from a misunderstanding of what the conjecture actually is.  Nevertheless there are a number of reasonable grounds by which one might be tempted to dismiss the  idea as crazy.  Fundamental questions about the approach include:
\begin{itemize}
\item Would effective chiral restoration actually look like the data?  Would effective chiral restoration necessarily lead to a massless fermions?

\item Is the idea even well posed?  Can it be falsified?

\item Can spontaneous symmetry breaking turn off smoothly with the mass of the state?

\item Does effective chiral restoration lead to an unnatural decoupling from pions?
\end{itemize}

As will be discussed in this section there are sensible answers to all of these questions.  Thus, while the idea may well turn out to be wrong, it is not crazy---one cannot rule it out on simple theoretical grounds.

\subsection{Does effective chiral restoration necessarily lead to massless fermions?}

The usual paradigm for chiral symmetry is for a Dirac particle:
\begin{equation}
\overline{q} \left ( i \partial_\mu \gamma_\mu + m \right)q
\end{equation}
Now, such a system is invariant under the chiral rotation $q \rightarrow e^{i \gamma_5 \theta} q$ only if the mass term is zero.  QCD at the quark level falls into this basic paradigm---the quarks are Dirac particles.

However, this does not mean that at the hadronic level chiral symmetry means massless fermions.  The baryons are most emphatically not Dirac fermions and need not be massless.    The key thing is that QCD in the massless limit is chirally symmetric---meaning that the theory is invariant under vector and axial rotations.  These are encoded in a Lie algebra---a set of commutation relations:
\begin{eqnarray*}
\left [V_a,V_b\right ] &= &i \epsilon_{a b c} V_c \\
\left[A_a,V_b\right] &= & i \epsilon_{a b c} A_c \\
\left[A_a,A_b\right] &= & i \epsilon_{a b c} V_c \; .
\end{eqnarray*}
Now the question is whether fermions can respect these relations without being massless.

As it happens, it has been known for a {\it very} long time that realizations of chiral symmetry exist in which all fermions are massive. Indeed, Benjamin Lee\cite{LEE} constructed such a realization in 1972---a year before QCD.  It is interesting to note that he dismissed such a construction as being ``physically uninteresting''.  The reason, of course, was that such construction had chiral multiplets for all baryons---whereas in nature baryons were found without parity partners.   As it happens a this idea was revived in an attempt to make realistic models of baryons\cite{DETAR,Mirror1,Mirror2,JIDO,TIT}.  The new ingredient was the inclusion of some form of spontaneous chiral symmetry breaking to split the chiral partners.  These models typically go under the name of ``mirror'' symmetry models.  The purpose of the models was typically to describe the low-lying baryons, such as the nucleon, and the $N^*(1535)$ (in which ``would be'' multiplets are badly split) rather than to explore the question of whether the high-lying states can naturally have a multiplet structure of nearly degenerate multiplets.  Still the models illustrate some of the key issues.

A simple version of the model is
\begin{eqnarray}
{\cal L} & = & \overline{N}_1 \left( i \partial_\mu \gamma_\mu + g_1\left (\sigma +\vec{\tau}\cdot \vec{\pi}  \right) \right ) N_1 \nonumber \\
 & + & \overline{N}_2 \left( i \partial_\mu \gamma_\mu + g_2\left (\sigma +\vec{\tau}\cdot \vec{\pi}  \right) \right ) N_2 \nonumber \\
 &{}& - m_0 \left (\overline{N}_1 \gamma_5 N_2 - \overline{N}_1 \gamma_5 N_2 \right ) + {\cal L}_{\rm meson}
 \end{eqnarray}
where ${\cal L}_{\rm meson}$ leads to a chiral symmetry breaking with $\sigma$ acquiring a vacuum expectation value of $\sigma_0$.
Note that the model is invariant under chiral symmetry provided that the axial transformation goes according to
\begin{eqnarray}
i[Q^a_R,N_{1 R}] &=& -i \tau^a N_{1 R} \nonumber \\
i[Q^a_L,N_{1 L}]& =& -i \tau^a N_{1 L} \nonumber \\
i[Q^a_R,N_{2 L}]&= &-i \tau^a N_{2 L} \nonumber \\
i[Q^a_L,N_{2 R}]& =& -i \tau^a N_{2 R}\; .
\end{eqnarray}
This is called a ``mirror'' realization since $N_2$ transforms in a ``mirrored'' way--its left-handed part transforms under the right-handed chiral transformations.

The model is typically treated in a mean field level.  In that case, it is easy to diagonalize the mass matrix.  One gets two states---one of positive parity and one of negative parity with masses given by:
\begin{equation}
m_\pm = \frac{1}{2} \left (   \sqrt{(g_1+g_2)^2 \sigma_0^2 + 4 m_0^2} \pm (g_1-g_2) \sigma_0   \right ) \; .
\label{mirror}\end{equation}

A couple of obvious comments:  Whatever the other virtues and vices the models have they are useful in illustrating what is needed in ``effective chiral restoration''.  Firstly, in the absence of chiral symmetry breaking $\sigma_0=0$ and the $m_+ =m_0$.  This is important since it demonstrates that a chirally symmetric model can have fermions  which are massive.  This ultimately proves the point noted earlier---nothing about chiral restoration implies massless fermions.  On the other hand, it is also clear that effective chiral restoration does {\it not} correspond to setting $\sigma_0$ to zero. $\sigma_0=f_\pi$ is a property fixed by the vacuum; vacuum properties are not altered as one goes up in the spectrum.  To get a nearly degenerate multiplet one needs the states to be essentially independent of the dynamics of effective chiral restoration.  In the context of this model this means the coupling to $\sigma_0$  must be small:
\begin{eqnarray}
g_1 \sigma_0  & \ll &  m_0 \nonumber \\
g_2 \sigma_0  & \ll &  m_0 \; .
\end{eqnarray}
It is easy to see from Eq.~(\ref{mirror}) that this does indeed lead to a nearly degenerate chiral multiplet.  Thus, in the context of this class of models the notion of effective chiral restoration is simply that as masses of baryons increase, the coupling $g_1$ and $g_2$ become small.

The general issue of the coupling of chirally restored states to pions will be discussed in detail later.  The coupling of pions to the baryons in this model will be of interest as in the illustrative example.  The couplings are given by
\begin{eqnarray}
g_{\pi N_+ N_+} & = &g_1 \cos^2(\theta) + g_2 \sin^2(\theta)  \nonumber \\
g_{\pi N_- N_-} & = & - \left( g_2 \cos^2(\theta) + g_1 \sin^2(\theta) \right ) \nonumber \\
g_{\pi N_+ N_-} & = &\frac{ g_2- g_1}{2} \sin(2 \theta)  \nonumber \\
\tan(2 \theta)&=& \frac{2 m_0}{\sigma_0 (g_1+g_2)}
\end{eqnarray}
Now consider the fact that effective chiral restoration require small values for $g_1$ and $g_2$ (perfect restoration would have them vanish). Thus, in the regime of effective chiral restoration the coupling to pions is small (zero for perfect restoration).  As will be discussed later, this feature of weak coupling to pions in the regime of effective chiral restoration appears to be generic.

A few words of caution about this class of models:  firstly, these are not systematic effective field theories---they are not based on any consistent power counting rules.  Instead they are models.  Moreover, the treatment of the models at the mean-field level is entire {\it ad hoc}.  Nothing in QCD suggests that this is a self-consistent approximation.  Note that from the perspective of the $1/N_c$ expansion, $g_1$ and $g_2$ both scale as $N_c^{1/2}$.  Thus, generically they are not small and thus there is {\it a priori} argument as to why loop effects need be small.

Whatever  one thinks of these "mirror symmetry" models, they clearly show by explicit construction that "effective restoration" in the sense of insensitivity to dynamics of chiral symmetry breaking does not automatically lead to massless fermions.  At least in this sense, the approach has passed a sanity check---it is not obviously nuts.  At least not on these grounds.

\subsection{Is the idea of effective chiral restoration well posed?}

The philosopher of science, Karl Popper, has argued that in order for an idea to be scientific, it must be falsifiable.  This raises an obvious question with regard to the notion of effective chiral restoration---is it scientific in a Popperian sense?  An obvious problem with the idea of effective chiral restoration is that the idea is somewhat qualitative.  As noted in the introduction, the question of whether or not one has  seen a chiral multiplet may depend on the number of glasses of good Austrian beer one has consumed.  See Fig.~\ref{beer}.  Part of the problem is that the notion of a hadronic resonance itself is intrinsically problematic to quantify.

\begin{figure}
\centering
\includegraphics[width=3.in]{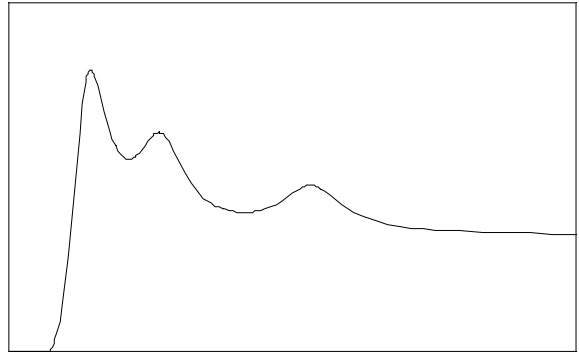}
\caption{A cartoon of a spectral function.  Note it has clear resonant structures as well as a continuum.}
\label{rho}
\end{figure}

The natural object to study is the spectral function, $\rho(s)$.  The spectral function is defined in terms of some current with the quantum numbers of interest and is proportional to the square of the amplitude that the current produces a state of invariant mass $\sqrt{s}$.  Correlation functions are fully determined by the spectral function through dispersion relations.  In principle $\rho(s)$ is computable to some level of accuracy via the lattice but as note earlier this is generally exponentially difficult.  Even if one can compute these, there is a further problem; the spectral function includes both continuum and resonant parts.  In Fig.~\ref{rho} a cartoon of a spectral function is given.  The problem here is that given the spectral function, how similar do peaks have to be before being declared ``nearly degenerate''? In a strict mathematical sense this is not well posed.

One might hope to learn qualitative features of the spectral function, even if one cannot fully characterize it.  At large space-like $Q^2$ the correlation functions can be computed perturbatively with power law corrections due to condensates.  A natural way for this to happen is for the spectral functions themselves to become increasing accurately described by perturbation theory.  Does this actually happen?  Well, in the case of the vector correlator, the spectral function is directly measurable via electron-positron scattering and at high momentum is indeed given to good approximation by perturbation theory.  Moreover, the phenomenologically successful QCD sum rule approach\cite{SVZ} is based precisely on this happening.

However, we know that perturbation theory cannot break chiral symmetry.  Thus to the extent that it is indeed true that QCD spectral functions at very high momentum are given perturbatively, it must be true that spectral functions for operators connected by chiral transformations (such as the current $\overline{q} i \gamma_5 \tau_a q$ with pion quantum numbers and $\overline{q} q$ with $\sigma$ quantum numbers) become identical at large $s$---{\it i.e.}, the states are degenerate.

Is this ``effective chiral restoration''?  The answer is no.  Note that by the time one is high enough up in the spectrum for the spectral strength to be perturbative, by construction there are no resonances left to observe.  A key question is whether there are still discernable resonant structures by the time  one is high enough in the spectrum that the effects of spontaneous symmetry breaking are small.  We do not know this {\it a priori}.  One might argue that the phenomenological evidence for effective chiral symmetry breaking suggests that it does, but the evidence seems suggestive rather than compelling.  From the point of view of theory we have no good approach to assess this question in general.

Fortunately, there is a scenario where the issue can be assessed cleanly, namely mesons in the large $N_c$ limit.  In this case, the meson spectrum remains discrete all the way up.  Nevertheless, the perturbative arguments for the correlator remain valid.  This appears to suggest that the spectra for chiral partners should become degenerate.  Shifman considered the meson spectrum at large $N_c$ and showed how chiral restoration could be approached\cite{shifm}; a similar analysis was done by Beane\cite{beane}.  Of course, this is a large $N_c$ argument and tells us nothing directly about the $N_c=3$ world.  Moreover, it tells us nothing about baryons.  It nevertheless is a proof of principle that hadrons can exist as discernable resonances high enough in the spectrum for effective chiral restoration to take place.

Unfortunately, it remains an open question whether effective chiral restoration {\it does} take place even for mesons at large $N_c$.  Note  that the previous argument that the spectra for operators related by chiral transformations becoming identical at large mass depended on the spectral functions becoming perturbative at large $N_c$.  However all we {\it know} is the correlation functions become perturbative at large space-like momenta.  As noted by in Refs. \cite{GP,cata} this does not in general require the spectra to be degenerate at large $N_c$.  If states are  offset from each other power law corrections are introduced to the correlator---but we know that power law corrections are consistent with the operator product expansion.  An alternative way to state the same thing---we only {\it know} that the spectral functions of two operators related by chiral transformations are the same if they are perturbative.  While there is some evidence empirically that spectral functions do become perturbative there is no general proof.  Moreover, we know that at large $N_c$ they do not: at large $N_c$ the spectrum is discrete all the way up while the perturbative spectrum is continuous.

Thus, while this line of inquiry can show that discrete levels consistent with effective chiral restoration {\it can} exist at large $N_c$, they do not prove that it must.   It is precisely because one cannot actually show this; that one cannot argue that effective chirally restoration trivially follows from the scales in the problem.  On the other hand, the large $N_c$ limit for mesons is important intellectually.  Since we know that at large $N_c$ mesons exist as narrow states all the way up in the spectrum,  the question of whether   masses and residues in the spectral functions for two operators connected by chiral transformations do approach each other for large masses can be posed sharply.  If they do not, the the idea is wrong.  In principle, this question can be answered; the idea of effective restoration is falsifiable.

While the idea is falsifiable in principle, in practice it is extremely hard.  The only known way at present to compute the spectra for high-lying states from QCD is via the lattice which, as we have observed before, is presently well beyond our computational means.

We should also note that at a more pedestrian level one can take steps which could falsify the idea of effective chiral restoration.  Note that there are missing levels in the empirical assignments of states.  If one were to more-or-less rule them out experimentally one could make  the notion that effective chiral restoration is quite unattractive.

\subsection{Can spontaneous symmetry breaking turn off smoothly with the mass of the state?}

One might worry that the underlying idea of a symmetry ``turns off'' smoothly as one goes higher in the spectrum.  In point of fact there is nothing problematic about this.  A simple model\cite{CG2} illustrates how this can happen.  It is quite plausible that the issue does not really depend on whether the symmetry breaking is spontaneous or explicit, but whether as one goes to higher excitation the effect of the symmetry breaking can smoothly turn off.

Consider the following simple two-dimensional quantum mechanical model.
It is a perturbed two-dimensional harmonic
oscillator.  The unperturbed system has a
 Hamiltonian which is invariant under
$U(2) = SU(2)\times U(1)$ transformations. This
symmetry causes a characteristic pattern of degeneracies for the unperturbed system.  The energy level of a general axially symmetric system depends on two quantum numbers; a principle quantum number, $N$, and  a rotational quantum number, $m$. In general, the energy for different values of $N$ and $m$ differ---although for a time-reversal invariant system $E(N,m)=E(N,-m)$.  However, in the unperturbed harmonic oscillator the energy levels are easily computable and are given by
\begin{eqnarray}
E_{N, m} \, & = &\, \omega ( N \, + \, 1 )\\
  m \, & = &
\, N, N-2, \, \cdots \, , -(N-2) , -N \; ,
\label{hoeigen}\end{eqnarray}
where $\omega$ is the frequency of the oscillator.  Note that the spectrum is highly degenerate---the energy depends on $N$ but not $m$---and a degeneracy of $N+1$ levels results.  The degeneracy is a consequence of the symmetry.

Consider what happens if we break this symmetry by adding to the Hamiltonian a time-reversal invariant perturbation which breaks the $SU(2)$ symmetry (while preserving the $U(1)$).  In general, this will split the degeneracy leaving behind only doublets connected by time-reversal $E(N,m)=E(N,-m)$.  If the symmetry breaking is strong, there will be no obvious remnant of the $N+1$-plet in the spectrum.  Consider, for example, the following perturbation:
\begin{equation}
V_{\rm SB} \, = \, A \, \theta (r - R) \;
\label{vsb}
 \end{equation}
\noindent
 $A$ and $R$ are parameters and $\theta$ is the step
function.    It should be obvious that
$V_{\rm SB}$ is not symmetric under the
$SU(2)$ transformation. For sufficiently large $A$ and $R$ the effect of the perturbation will be large and the $N+1$-plet structure will completely be lost.

\begin{figure*}
\centering
\includegraphics[width=5.5in]{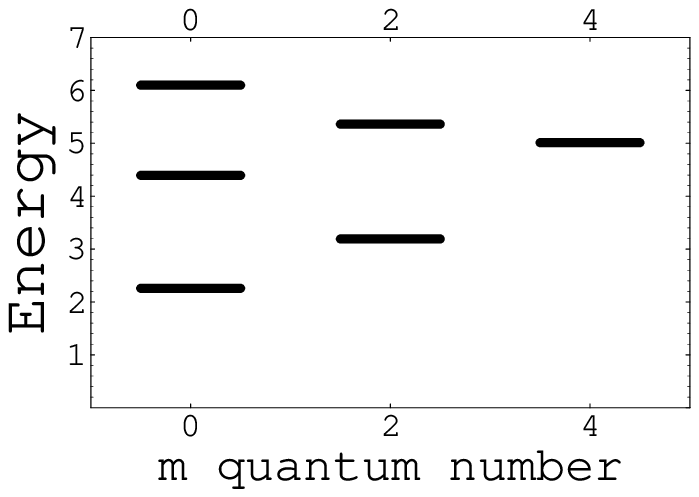}\\
\includegraphics[width=5.5in]{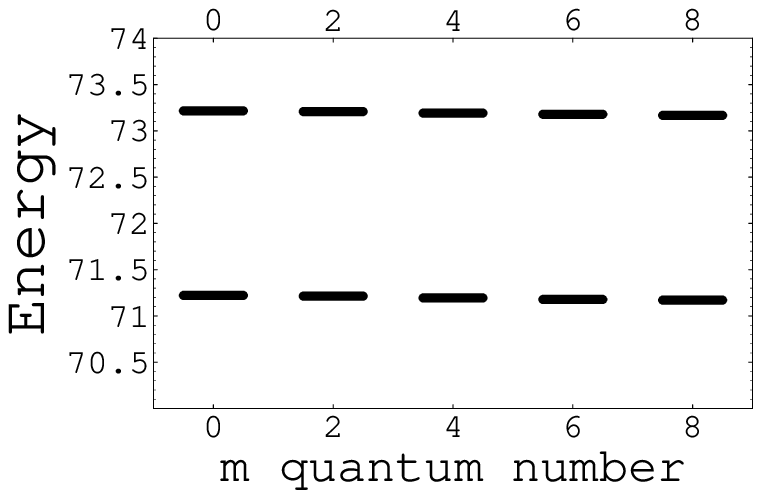}
\caption{Low-lying (top panel) and highly-lying (bottom panel)
spectra of the two-dimensional harmonic oscillator with the
$SU(2)$-breaking term.}
\label{to}
\end{figure*}
This model has been solved
numerically for the energy eigenstates for the case of $A=4$ and $R=1$
(in  dimensionless units  with $\omega=1$).  Some representative levels are plotted in Figs. \ref{to}.  As advertised, the perturbation badly splits the ``would be'' $N+1$-plet for the low-lying levels.  The interesting point for the present context is what happens at high excitations.  We have plotted energies from 70 to 74 in dimensionless units and to extremely high accuracy states of the same $m$ are degenerate.  That is, the $SU(2)$ symmetry which was badly broken by the perturbation is effectively restored high in the spectrum.

One can easily understand why the $SU(2)$ symmetry  is effectively restored at high excitation by thinking about the nature of the wave functions.  This issue will not be pursued here as the mechanism of how the symmetry is effectively restored is obviously quite different than in the case of the spontaneous breaking of chiral symmetry (assuming it really is effectively restored).   What this simple model does, however, is to demonstrate via explicit construction the fact that a broken symmetry can become effectively restored high in the spectrum.  Thus one cannot rule out the conjecture on the grounds that it is somehow unnatural for symmetry breaking to smoothly turn off high in the spectrum.  That is to say, the conjecture of effective restoration of chiral symmetry may be wrong---but at least it is not crazy (at least on these grounds).

\subsection{Does effective chiral restoration lead to an unnatural decoupling from pions?}

In the mirror symmetry models it was seen that pions were uncoupled from the baryons if chiral symmetry was effectively restored.  While that was in the context of a particular model, there is an argument that it is a generic feature.  The argument goes something like this:  effective chiral restoration means the hadron is insensitive to the dynamics responsible for breaking chiral symmetry breaking---{\it e.g.}, the chiral condensate. Since the pion is essentially just a fluctuation of the chiral condensate, a hadron in the effectively restored regime ``doesn't know'' whether or not there is a pionic fluctuation of the condensate.  That is, the hadron would be decoupled from pions.
There are two issues here: (i) whether the heuristic argument that effective chiral restoration implies decoupling from pions is correct, and (ii) whether this is a phenomenological problem.

Let us address the first issue.  There are many reasons to believe that the argument is sound.  It is plausible enough on its own.  Moreover, it can be seen to occur in models.  It was shown to occur in the mirror nuclei models.  It can similarly be shown to occur in a wide variety of mean-field models.  (Mean-field plays an important role in that the mean-field approximation does not violate symmetries and symmetry plays an essential role here.  Most approximation schemes explicitly break symmetries).

\begin{figure*}
\centering
\includegraphics[width=4.0in]{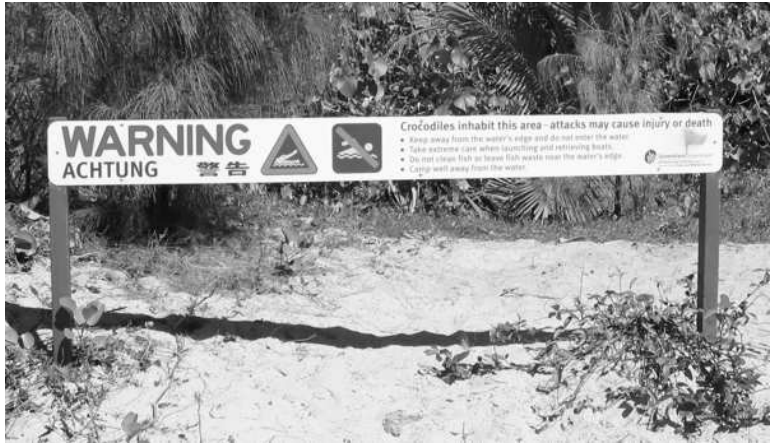}

\caption{Trusting models of confinement is like swimming with crocodiles.}\label{croc}
\end{figure*}

It can also be seen to occur in a quark-based model which explicitly includes chiral symmetry breaking and confinement\cite{WG1}.  Now, it is reassuring that confinement does not destroy the phenomenon.  At the same time, it is important to be very cautious in interpreting such a model.  The basic problem is that even after all  these years we really do not understand the confinement mechanism in QCD.  Indeed, I would argue we do not know whether the notion of a ``confinement mechanism'' is even well defined.  That being the case, it is hard to know whether confinement as implemented in any simple model has {\it anything} to do with confinement in QCD.  I have suggested that trusting models of confinement is like swimming in crocodile-infested waters. See Fig.~\ref{croc}. (It is noteworthy that one of the authors of ref.~\cite{WG1} was observed swimming in the Coral Sea in front of the sign in Fig.~\ref{croc}.)  With these warnings in mind, it is still encouraging that the model---which implements confinement via a linear rising static interaction in a Bethe-Salpeter amplitude {\it does} see both effective chiral restoration and the decoupling of pions.

Is there a more general and formal way to see it?  The answer is ``yes'', but only in a limited sense.  The MIT group\cite{jaffeDecoup} has shown that to the extent that one has well-defined hadronic states, a generalized Goldberger-Trieman relation provides some insights. For simplicity, consider a parity-doublet of baryons ({\it i.e.}, the $(1/2,0) + (0,1/2)$ chiral representation), although an  analogous argument can be constructed for other representations as well. One starts by writing the matrix elements of the axial current between the positive and negative parity hadron and parameterize it in its most general form:
\begin{eqnarray}
\label{current}
\langle B_{+} |j_{5\ \mu}^{a}|B_{-} \rangle & =  & \bar u(p,s)  \Gamma(q^2) t^{a} u(p',s')\nonumber \\
\Gamma(q^2) \equiv  \gamma_{\mu}g_{A}(q^{2})&  +  &q_{\mu}g_{P}(q^{2})
+i\sigma_{\mu\nu}q^{\nu}g_{M}(q^{2}) \, \nonumber \\
q^{\mu} &\equiv &p^{\mu}-p'^{\mu}
\end{eqnarray}
 The axial form factor is $g_A(q^2)$; similarly $g_P(q^2)$ is the induced pseudoscalar form factor.
The axial current is conserved, implying that  $q^{\mu}\langle B_{+}|j_{5\
\mu}^{a}|B_{-}\rangle =0$.  This in turn implies that
\begin{equation}
\label{gt}
 \Delta m g_{A}(\Delta m^{2})+\Delta m^{2}g_{P}(\Delta m^{2})=0
\end{equation}
with $\Delta m=m_{+}-m_{-}$.

The induced pseudoscalar form factor $g_{P}(q^{2})$ has a pion pole.  In the exact chiral limit, Goldstone's theorem tells us that it is at $q^{2}=0$.  Its residue is given by
$f_{\pi}g_{\pi B_{+}B_{-}}$,  where $g_{\pi B_{+}B_{-}}$ is  the
$B_{+}B_{-}\pi$ coupling constant (which is for an s-wave coupling due to the parities of the two states).  Now suppose that $\Delta m$ happens to be small compared to typical hadronic scales.  In this case we have
\begin{equation}
\Delta m g_{A}+f_{\pi}g_{\pi B_{+}B_{-}}+{\cal O}(\Delta m^{2})=0
\label{fullgt}
\end{equation}
where $g_{A}=g_{A}(0)$.

Let us now consider what happens if effective chiral restoration occurs.  In that case $\Delta m=0$ and Eq.~(\ref{fullgt}) implies that the pion coupling vanishes.
%\begin{figure}
%\centering
%\includegraphics[width=3.in]{selfen.eps}
%\caption{Self-energy contributions for two baryons of opposite parities in which each baryon breaks up into nucleon-pion pairs.}
%\label{selfen1}
%\end{figure}

At first blush, it looks like the preceding argument is sufficient to prove that pions must decouple if effective chiral restoration occurs.  However, this is not quite true.  In the first place, strictly speaking, the matrix elements in question are only defined for single-particle states.  However, the hadrons in question are resonances, not single-particle states.  One might hope that this problem is not too serious if the baryon resonances are reasonably narrow, but the issue is a cause for concern.

There is a more serious problem.  This argument only shows that the pionic coupling between the nearly degenerate states is small---zero in the exactly restored limit.  It does not tell us that pions cannot couple two hadrons which are not nearly degenerate---{\it i.e.}, which are not part of the same chiral multiplet.  To see why, let us go back to Eq.~(\ref{gt}), but consider the case where the matrix element is for two hadrons with different masses.  (This  could happen if the two baryons are each in the regime of effective chiral restoration but are part of different chiral multiplets, or if one is in the chirally restored phase and the other not.)   In that case Eq.~(\ref{gt}) does {\it not} automatically imply that the pion couplings are zero.   Of course, it is perfectly consistent with them being small as well---the two form factors themselves can vanish.

To summarize, the argument based on a generalized Goldberger-Trieman relation shows that assuming effective chiral restoration occurs and that states are narrow, pionic couplings within a chiral multiplet are small (zero in the restoration is exact), but tells us nothing, one way or the other, about pionic coupling {\it between} multiplets.

However, apart from the qualitative argument given earlier that the pion is a fluctuation of the chiral condensate, there is another general---if hand waving---argument as to why pionic couplings to states of different masses should also be small in the chirally restored phase.  Consider the contribution to the mass of two different baryons due to a self-energy contribution in %which the baryon breaks into a pion-nucleon pair which subsequently annihilate as
 %in Fig.~\ref{selfen1}
which the baryon breaks into a pion-nucleon pair which subsequently annihilate.

It is worth recalling that the self-energy is not generically small.  In terms of $N_c$ counting they make a leading order ({\it i.e.}, order $N_c^1$) contribution to the nucleon mass. Now suppose that the baryons are in the regime of effective chiral restoration and are members of the same chiral multiplet.  In that case their masses should be nearly degenerate.  On the other hand, the self-energy contributions are of generically very different forms---the differing parities ensure this.  Thus one  generically does not expect the self-energies to be the same for the two channels.  For the two baryons to be degenerate, one of two things needs to happen: either a large conspiracy of unknown origins in which the differing self-energies are compensated by complementary differences elsewhere, or the self-energy contributions for both happen to be small.  It is difficult to envision how such a conspiracy could occur so the natural way for the states to remain degenerate is if the self-energies are forced to be small.  This in turn happens if the coupling of the baryon into a nucleon plus pion is small.

Actually the preceding argument is more general than just suggesting that the coupling of hadrons in the regime of effective chiral restoration to pions is small: it suggests that couplings to {\it all} hadrons which are {\it not} effectively restored are small.  The preceding argument can be recast using any hadron in the self-energy.    It is noteworthy that for the case of coupling to other chirally restored hadrons the argument does {\it not} imply small couplings.  
%To see why consider Fig.~\ref{selfen2}.  In  these diagrams contributions to the self-energy have each of the two baryons in an effective chiral multiplet each  breaking up into an excited nucleon and a meson.  Note that the excited nucleon and the meson in the second diagram are related to their analogs in the first by the same chiral transformation relating the baryons.  This in turn implies that the structure of the baryon-nucleon-meson vertices in the two graphs are the same and thus the self energy contributions are the same.  Therefore, degeneracy can be maintained without relying on an unnatural conspiracy.
To see why consider contributions to the self-energy have each of the two baryons in an effective chiral multiplet each  breaking up into a highly excited nucleon and a meson.  Note that in this case the excited nucleon and the meson in one self energy are related to their analogs in the other by the same chiral transformation which  relates the baryons.  This in turn implies that the structure of the baryon-nucleon-meson vertices in the two graphs are the same and thus the self energy contributions are the same.  Therefore, degeneracy can be maintained without relying on an unnatural conspiracy.

The upshot of this general argument is that one naturally expects that if effective chiral restoration were exact, a chirally restored hadron could only decay into other chirally restored hadrons.

To summarize the overall situation:  there is a general plausibility argument based on sensitivity to the dynamics of chiral symmetry breaking that hadrons in the regime of effective chiral restoration are decoupled from pions and this behavior is seen in simple models.  There is a formal argument showing that to the extent the hadrons are well-defined states, pionic couplings {\it within a multiplet} must vanish if chiral restoration is exact.  There is also a more general plausibility argument suggesting that if effective chiral restoration were exact, a chirally restored hadron could only decay into other chirally restored hadrons.

The weight of these arguments make it very plausible that effective chiral restoration, if it occurs, is accompanied by a decoupling from pions.  Indeed, this has been taken to be well established in refs. \cite{Cdec,MITrev} and \cite{jaffeDecoup}.  In this context, it is important to recall that while the plausibility arguments are quite strong, they do not rise to the level of a mathematical theorem and it remains logically possible that effective chiral restoration does not necessarily lead to decoupling from the pions.  However, for the remainder of these lectures, I will implicitly assume that the decoupling is correct.

%\begin{figure}
%\centering
%\includegraphics[width=2in.]{selfen2.eps}
%\caption{Self energy contributions for two baryons of opposite parities in which each baryon breaks up into an excited nucleon-meson pair.  The primes  indicate another member for the same chiral multiplet as the first and connected by a particular chiral transformation.}
%\label{selfen2}
%\end{figure}

Suppose, that decoupling is well established.  Should this be viewed as a bug---or a feature?  It has been suggested that the decoupling of pions from these high-lying hadrons is quite unnatural and constitutes a strong argument against the notion of effective chiral restoration.   Of course, it would be very unnatural if the coupling to pions were, in fact, zero.  Indeed, if the argument given above is true that effectively restored hadrons only couple to other effectively restored hadrons, it might seem that these highly-excited states can never decay.  However, it is important to recall that the notion of approximate chiral restoration is  {\it approximate}; thus one expects the coupling to pions for these states is weak (rather than vanishing) and presumably becomes progressively weaker as one goes up in the spectrum.   As it happens such a behavior may help solve one of the great puzzles in hadronic physics.

The puzzle: The width of a well-defined hadron depends on both the strength of the coupling to the decay channel and also on the available phase space.  Now the phase space grows quite rapidly with the mass of the resonance.  This leads to the natural expectation that hadron masses characteristically ought to grow with mass.  From this perspective, the fact that hadrons made of light quarks are hard to discern much about 2 GeV is simply because they get too wide to be seen. In fact, this is not really the case.  Hadron masses do not seem to grow characteristically with mass except, perhaps, quite slowly.

Consider for example the isoscalar mesons  (from the PDG):

\begin{center}

{\bf state~~~~~~~~~~~~~~~~~~~~~~~~~~~width}\\
\medskip
~~$f_0$(600)~~~~~~~~~~~~~~~~~~~~~~~~~$\Gamma \sim$ 800 MeV\\
~$f_0$(980)~~~~~~~~~~~~~~~~~~~~~~~~~$\Gamma \sim$ ~70 MeV\\
~$f_2$(1270)~~~~~~~~~~~~~~~~~~~~~~~~$\Gamma \sim$ 185 MeV\\
~$f_1$(1285)~~~~~~~~~~~~~~~~~~~~~~~~$\Gamma \sim$ ~25 MeV\\
~$f_0$(1370)~~~~~~~~~~~~~~~~~~~~~~~~$\Gamma \sim$ 350 MeV\\
~$f_1$(1420)~~~~~~~~~~~~~~~~~~~~~~~~$\Gamma \sim$ ~55 MeV\\
~$f_0$(1500)~~~~~~~~~~~~~~~~~~~~~~~~$\Gamma \sim$ 110 MeV\\
~$f_0$(1710)~~~~~~~~~~~~~~~~~~~~~~~~$\Gamma \sim$ 135 MeV\\
~$f_2$(1950)~~~~~~~~~~~~~~~~~~~~~~~~$\Gamma \sim$ 470 MeV\\
~$f_2$(2010)~~~~~~~~~~~~~~~~~~~~~~~~$\Gamma \sim$ 200 MeV\\
~$f_4$(2050)~~~~~~~~~~~~~~~~~~~~~~~~$\Gamma \sim$ 240 MeV\\
~$f_2$(2300)~~~~~~~~~~~~~~~~~~~~~~~~$\Gamma \sim$ 150 MeV\\
~$f_2$(2340)~~~~~~~~~~~~~~~~~~~~~~~~$\Gamma \sim$ 320 MeV\\
\end{center}

It is noteworthy that the widest state is the lowest.  It is arguable, of course, the the $f_0$(600) is not a ``real'' meson.  Still it is clear that many mesons remain rather narrow despite having very large phase space for decay.

Now one possible explanation for this is simply selection bias.  Suppose it were the case that there were very many highly-excited mesons, most of which are too wide to discern.  In such a case, the fact that the ones we see at high mass are relatively narrow may reflect the fact that those in the Particle Data Book are narrow because we would not have seen them if they were not.  There is an alternative possibility, namely that the states remain narrow simply because the gain in phase space as the mass increases is compensated for by a diminution of the coupling to channels containing two (or more) low mass particles such as the two-pion channel.

One natural way for this to happen is if effective chiral symmetry restoration becomes increasingly exact as the mass increases leading to particles which are increasingly weakly coupled to pions.

Is this what is happening?  There is one interesting bit of evidence suggesting that this scenario is correct, namely a systematic study of the decay of baryons into the pion-nucleon channel\cite{Cdec}.  If the basic scenario is correct, one would expect to find that high-lying states are in the form of doublets {\it and} the coupling to the pion-nucleon channel is small and decreasing with mass.  In fact, an extraction of the coupling for nucleon states with masses of 1440 MeV or larger indicates that {\it almost} all of the states have plausible chiral partners and that the couplings to the pion-nucleon channel are very small.  One can normalize the coupling against the usual pion-nucleon-nucleon coupling.  The natural thing to look at  is the square of the ratio of the couplings and the largest value it has is .15 ----and that is for the low-lying $N^*$(1440).  Higher-lying states have {\it very} small couplings.

What makes this data interesting is that there is only one case where there is not a plausible chiral partner and that is the $N^*$(1520). This state uniquely has a large coupling to the pion-nucleon channel---the  square of the ratio to the usual pion-nucleon-nucleon coupling is 2.5.  This suggests that having a small coupling to the pion nucleon channel appears to be correlated with the particle not being part of an effectively restored chiral multiplet.  Of course, as there is only one such example, one cannot claim a general pattern.  Still it is encouraging that things work out this way.

\section{Models}

As noted throughout these lectures the question of whether the scenario of effective chiral restoration actually occurs remains open.  However, the question of whether the scenario is {\it possible}---{\it i.e.}, whether it can be ruled out from general principles of chiral symmetry and known properties of QCD---can be addressed.  A natural way to do this is via models.  There have been a number of models which have been studied to address these issues.  I will not discuss them  all in any detail here, but rather I  summarize the overall situation and then discuss one illustrative model.

One expects {\it any} model in which hadrons exist as narrow states arbitrarily high in the spectrum and which has a single mechanism of spontaneous chiral symmetry ({\it eg.}, coupling to the chiral condensate) will illustrate the phenomenon provided the coupling of the hadrons to the mechanism of spontaneous symmetry breaking weakens with excitation energy. A number of models of this sort have been constructed and they do indeed illustrate the issue.

\begin{figure*}
\begin{center}
\includegraphics*[width=6in]{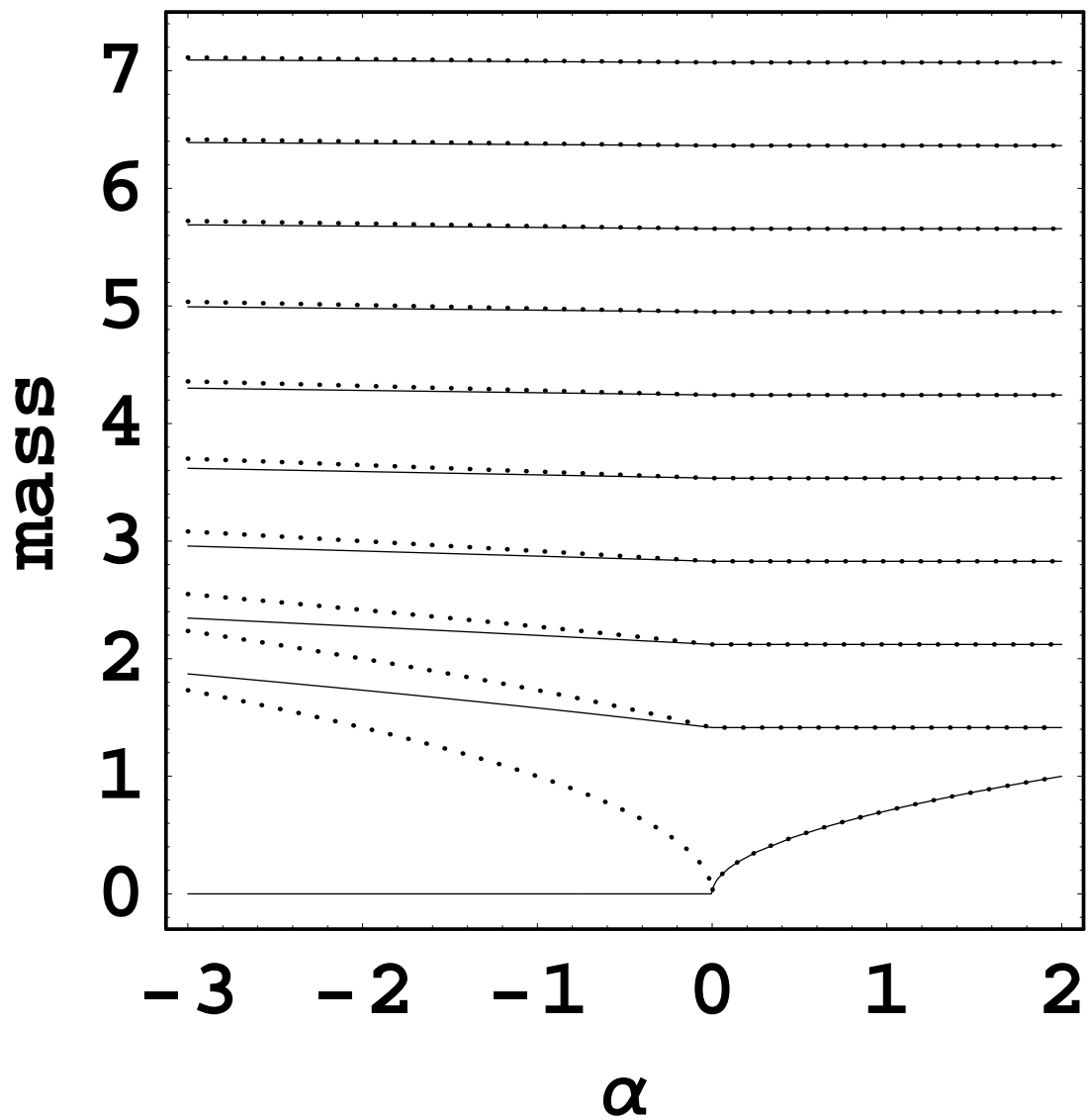}
\end{center}
\caption{The mass spectrum of the model in Eq.~(\ref{L}) illustrates the phenomenon of effective chiral restoration. \label{toy}}
 \end{figure*}

Figure \ref{toy} shows the spectrum of one such model.  The model\cite{CG3} is based on the linear sigma model\cite{GL} with  pion and $\sigma$ meson fields forming chiral partners.  To simulate the fact that large $N_c$ QCD has an infinite number of mesons, the model builds in an infinite number of such pairs.  The Lagrangian is given by
\begin{eqnarray}
{\cal L} &=& \sum_j \frac{1}{2} \left ( \partial^\mu  \sigma_j \,
\partial_\mu \sigma_j \, +
\partial^\mu  \vec{\pi}_j \cdot \partial_\mu \vec{\pi}_j  \right ) \nonumber\\
 &-& \frac{m_o^2}{2} \left ( \alpha (\sigma_1^2 + \vec{\pi}_1 \cdot
\vec{\pi}_1 )  +  \frac{g}{2 m_o^2} (\sigma_1^2 + \vec{\pi}_1
\cdot \vec{\pi}_1 )^2 \right ) \nonumber \\
& - &  \frac{m_o^2}{2}
\sum_{j=2}^\infty \, \left( j^2 (\sigma_j^2 \, + \vec{\pi_j} \cdot
\vec{\pi_j}) \right ) \nonumber \\
&+& \sum_{j=2}^\infty \,  \frac{g}{j \, m_o^2 \,} \left ( (\sigma_1
\sigma_j + \vec{\pi}_1\cdot \vec{\pi}_j)^2  \right ) \nonumber \\
&+& \sum_{j=2}^\infty \,  \frac{g}{j \, m_o^2 \,}
(\sigma_1^2  +
\vec{\pi}_1\cdot \vec{\pi}_1) \, ( \sigma_j^2 + \vec{\pi}_j\cdot
\vec{\pi}_j)   \label{L}
\end{eqnarray}
The model is treated in the classical limit---which is justified in the large $N_c$.  Spontaneous symmetry is encoded in the potential for the first pair.  The parameter $\alpha$ controls the symmetry breaking---it breaks for $\alpha <0$ and the larger the negative value $\alpha$ has the more the symmetry is broken.  The model builds in decreasing coupling to the condensate for increasing masses.  It clearly demonstrates the phenomenon of effective chiral restoration.

However, it is worth noting that models in which hadrons exist as narrow states arbitrarily high in the spectrum can also be constructed that do {\it not} exhibit effective chiral restoration.  The fact that models consistent with chiral symmetry and its spontaneous breaking can either have effective restoration or not shows that the phenomenon if true depends on detailed QCD dynamics and cannot be deduced entirely based on general arguments.

\section{Stringy Descriptions}

I have tried in these lectures to suggest that effective chiral restoration may be a possible explanation of the pattern of degeneracies seen in excited hadrons.  There are certainly hints in the data which suggest this scenario.  However, it is important to be cautious and see whether there may be other explanations which also describe the data.

An anecdote illustrates the need to remain open-minded with regard to alternative explanations even when one has a very reasonable explanation of the data. One of the great features of the Schladming Winter School is its location at the foot of Planai---an excellent mountain for skiing.  As it happens, my daughter Becky, was spending the year studying art in Florence and fortunately her art school had a week-long break exactly during the week of the Schladming Winter School. It was natural for her to join me in Schladming for the skiing.  Our first day, we arrived in the late afternoon and Becky went off to rent skis for the next day.  She brought the skis, boots and poles back to the apartment which she and I were sharing with my student Aleksey.  The next morning she gathered up the ski equipment, found me at the conference site (I had brought along my skis) and off we went up the Planai.

Aleksey, who had brought his skis from the US, did not join us skiing that day.  Now, as we started to ski Becky had a major problem with her boots---she could not get them adjusted---they felt awful and she had trouble skiing in them.  She eventually skied down to the rental shop where it was realized that they were two sizes too large.  The shop exchanged them for the right sized boots and she went back up the mountain.

Now a question arose---how could the shop have given her boots two sizes too large?  We formed a hypothesis fully consistent with the data we had.  In the United States women's sizes and men's sizes of the same number correspond to sizes which differ by two.  A size 8 men's shoe size is the same size as a size 10 women's.  Thus our explanation was this---relatively few Americans ski in Schladming so that when Becky gave her shoe size, the ski shop mistakenly converted an American's men's size to a European size resulting in a boot two sizes too large.  Presumably when trying it on she had not noticed just how large it was until trying to adjust to go skiing.  In any, event it seemed alls well that ends well.  With the now correct sized boots Becky had a great time skiing.

However, things got interesting two days later when Aleksey decided to ski.  He went to the front hall of our apartment where the ski equipment was kept and said ``where are my boots''.  When I pointed to a pair on the floor he said, ``those aren't my boots''.  Indeed he was correct---the boots on the floor had a ski shop rental number on them.

I will spare you the details of our adventures getting Aleksey's boots back from the ski shop.  What made the whole little incident possible was the fact that Aleksey's boots---though two sizes larger than Becky's---had a sole of exactly the same length and precisely fit her bindings.  There was no hint that something had gone wrong---other than Becky's boots being too large.  A second thing which made the situation possible was that the ski shop in accepting the boots back failed to notice that they lacked a rental number on them.  Now, the interesting point here is that our perfectly rational explanation for why Becky's boots were two sizes too large was completely wrong and in the absence of additional data we would not have known.

I bring this little story up to point out that just because the boots fit into the binding does not mean our original explanation was correct.  A possible analogy---just because the $N*$(1520), the state without a potential chiral partner, has a large coupling to the pion-nucleon and channel, and thus fits neatly into the scheme of effective chiral restoration, does not mean the scheme is correct.  It has been suggested that another explanation---stringy dynamics could explain the pattern of degeneracies.\cite{A}

\subsection{Hadronic strings}

It is worth reminding ourselves that modern string theory grew Phoenix-like from the ashes of a failed attempt to describe strong interactions as a string theory.  While the approach had various phenomenological successes, it was ultimately abandoned  as the fundamental theory of strong interactions.  Firstly it had
phenomenological problems (a pesky massless spin-2 meson and the like). Secondly there were problems of theoretical consistency ({\it eg.}, negative norm states and tachyons\cite{string}).  Finally, the emergence of QCD as a viable field theory for strong interactions ultimately removed string theory from consideration as an ultimate description of strong interactions.

However, it quickly became accepted lore that QCD becomes stringy for highly excited states (at least at large $N_c$).  There are some deep theoretical reasons to believe this.  In the first place, there are reasons to believe that confinement in QCD takes the form of an area law for the Wilson loop  at long distances (at least for the large $N_c$ limit where center symmetry is well defined).  This is easily understood if QCD (at large $N_c$) develops a flux tube.   Moreover lattice studies confirm this picture.  While this picture is only known to apply for static flux tubes it is reasonable to assume in the regime of sufficiently high energies, so that the flux tube is stretched to the point that it is much longer then it is wide---its dynamics will essentially be stringy.  Large $N_c$ plays a role in as much as it suppresses string breaking.  Such a picture has the virtue of only become effectively stringy for highly excited states.  This is of value in that the four-dimensional string theory itself has diseases at low momentum (such as massless spin-2 boson).  The hope is that corrections due to finite $N_c$ will yield quantitative rather than qualitative corrections to such a stringy picture for high-lying states.

The spectrum of hadrons in QCD gives another reason to believe QCD becomes stringy, at least at large $N_c$.  One consequence of any string description is the existence of a Hagedorn spectrum\cite{string,Hagedorn1,Hagedorn2}:
\begin{equation}
N(m) \sim \exp (m/T_H)
\end{equation}
where $N(m)$ is the number of hadrons with mass less then $m$ and $T_H$, the Hagedorn temperature, is a parameter.  In the context of QCD, $T_H$ represents  an  absolute upper bound for hadronic matter.  Empirical studies of QCD with $N_c=3$ suggest that the growth of the number of states is consistent with exponential growth\cite{Hagexp}.  It is not clear whether this empirical data---restricted as it is to relatively low mass states---is sensitive to the dynamics actually responsible for exponential growth.  At a theoretical level, there has been a recent {\it ab initio} derivation of a Hagedorn spectrum directly from large $N_c$ QCD, provided certain standard technical assumptions about the applicability of perturbation theory to correlation functions apply\cite{CohenHag}.

Apart from the Hagedorn spectrum, there is another aspect of hadron spectroscopy that supports the notion of QCD as effectively a string theory for high-lying hadrons: the existence of Regge trajectories.  These again are known to arise in string theories.  Empirically Regge behavior is well documented in hadron spectroscopy.  More generally, {\it simple} string theories lead to the expectation that the square of meson (and glueball) masses should grow linearly with an integer quantum number $n$; the maximum value of the spin also given by $n$ giving rise to Regge behavior.  Such a picture gives rise to high levels of degeneracy between hadrons of different quantum numbers.  This could be the origin of the pattern of degeneracies seen in the data.

\subsection{A Baconian approach redux}

\begin{figure*}.
\includegraphics[width=6in.]{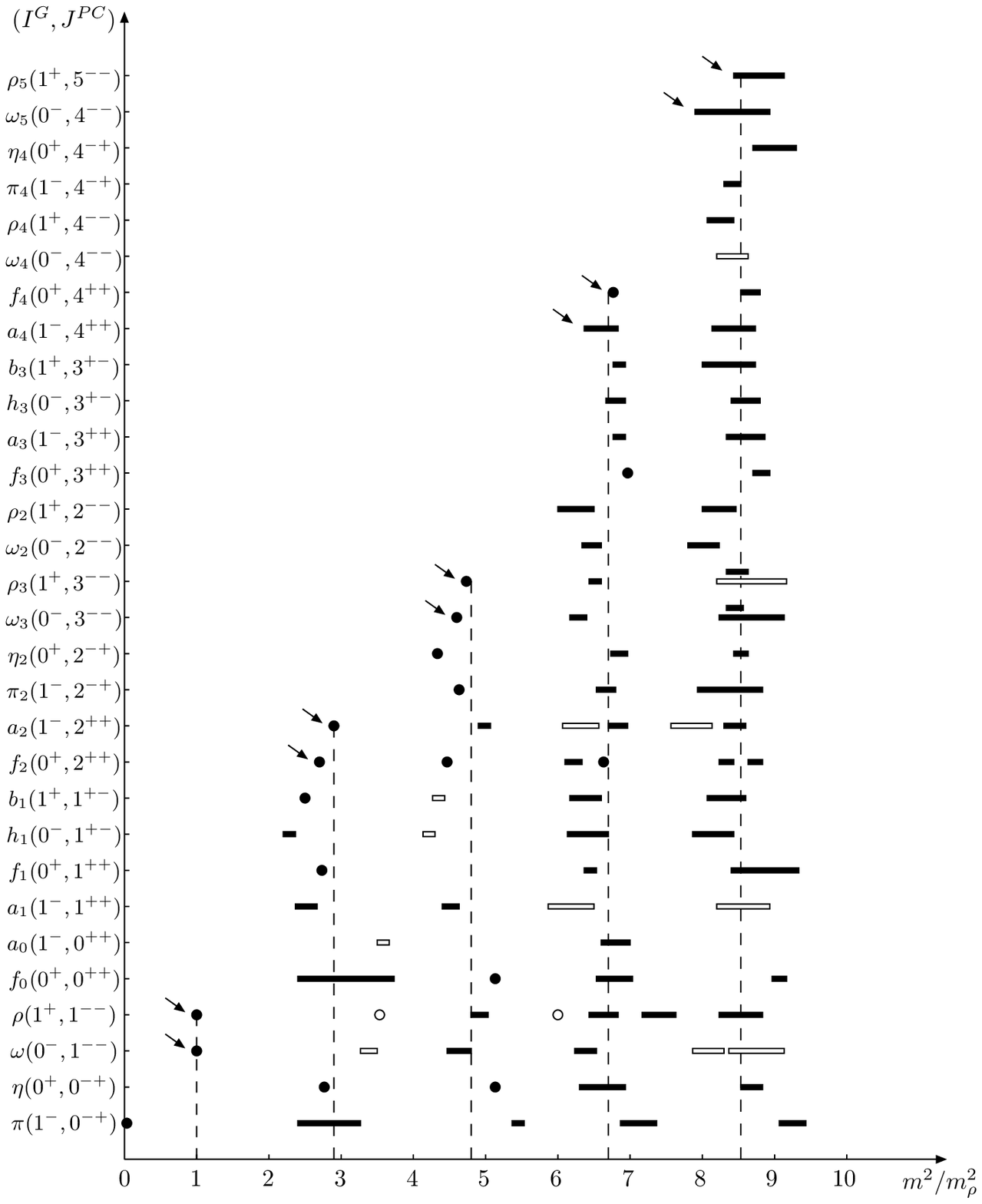}
\caption{Meson masses squared (in units of the $\rho$ meson mass) grouped into bands by $n$.  The arrows indicate levels without known parity partners.}
\label{stringfig}
\end{figure*}

Perhaps the best way to ask whether stringy dynamics is at work here is to look directly at the data.    For this purpose it is probably better to restrict our attention to mesons since stringy descriptions of baryons are intrinsically more complicated and potentially harder to interpret.  It is also sensible to look at data patterns for the square of meson masses rather than the masses themselves as the square of the masses have a simple stringy description.  Afonin compiled a table of such masses\cite{A} which is reproduced as Fig. \ref{stringfig}. It is highly instructive.

A few obvious statements about this data.  Firstly, there appears to be  discernible bands containing a number of mesons of relatively similar masses.  With a sufficient amount of good Austrian beer one could probably convince oneself  that this pattern is clear.  See Fig.~\ref{beer}. A second important feature is that these bands contain mesons of both positive and negative parities.   Finally, it should be clear that these bands are spaced at approximately equal distances.  This is characteristic of the behavior one expects if the data approximating a system is given by simple string-type models.

Before jumping to the conclusion that this data really does strongly support the string picture, a modest word of warning.  If the dynamics is in fact stringy, one would expect a Hagedorn-like spectrum.  This is possible only if there are approximate degeneracies of states in the bands {\it with the same quantum numbers}. In fact, the only way a Hagedorn spectrum can happen is if such degeneracies grow exponentially with the mass.  The fact that they are not seen in the data should make one a bit cautious in assigning a stringy character to the dynamics based solely on this data.  Now, of course, it is possible that  such degenerate states exist but have not been recognized experimentally: absence of proof is not the same as proof of absence.  Still, this is a cause for concern if one wishes to adopt a stringy description.

Suppose, for the sake of argument, that the stringy description is correct---does this mean that the description based on effective chiral restoration is wrong?  At first blush, the answer is no---one can easily imagine a situation in which the effective chiral multiplets are contained within a larger degeneracy associated with stringy dynamics.  However, there is a catch---if one looks at the data in Fig.~\ref{stringfig} one sees that the states at the top of each band (highest spin) have no parity partners.

Afonin\cite{A} has constructed a model motivated by a string picture which describes the data.   The model is very simple: the energies are given by
\begin{eqnarray}
m^2 & =& m_0^2 (l + n_\gamma +c ) \nonumber \\
l& =& 0, 1, 2, 3, \cdots n\; \; \; n_\gamma=0,1,2, \cdots \nonumber \\
J &=& l-1,1,l+1 \; \; \; P=(-1)^l
\end{eqnarray}
where $m_0$ and $c$ are parameters.  The model is basically that of an open string (with an offset $c$ due to low-lying dynamics) with  $J$, the spin of the state given by the orbital angular momentum of the string, $l$,  coupled to the spin of the quarks.  This model has the virtue that at least at the level of a cartoon, it fits the data.  A simple string picture has $m^2 =m_0^2 \, n $ with $l_{\rm max} =n$. Provided that one identifies $n=n_\gamma +l$, this is precisely the structure here (one  takes into the shift $c$ due to low-lying dynamics.  It is also clear why the highest J states in any band in this model does not have a parity partner.  The state with the maximum $J$ has $J_{\rm max}=l_{\rm max} +1$ and it has parity $(-1)^{l_{\rm max}}$.

While this does summarize the data well, there are some important problems with this simple picture.  It looks like an open string coupled to the spins of the quarks {\it provided that the quarks are nonrelativistic and have a weak spin-orbit coupling}.  However, such a description is an {\it assumption} beyond simple stringy dynamics.  Moreover, this assumption has major conceptual difficulties.

In the first place the model treats the angular momentum of the string and the quark spin as having good quantum numbers.  This is sensible only if the spin quark is (at least approximately) conserved---separately from the conservation of angular momentum in QCD. However, as far as we know only the total angular momentum in QCD is conserved.  Thus the model appears to be very different from QCD in a critical way associated with the assignments of the state's quantum numbers. Secondly, the assumption appears to be {\it inconsistent} with the string picture on which it is supposed to be based.  The string picture has the ends of the string looking like color sources moving at the speed of light.  If one identifies these with quarks then quarks are moving at the speed of light.  Such quarks are not exactly nonrelativistic---they are ultrarelativistic.  Since the model is based on adding a nonrelativistic spin, it is highly questionable.  The fact a string picture with quarks localized at the ends of the string has ultrarelativistic quarks suggests that the model {\it ought} to have an unbroken chiral symmetry.  After all, massless (that is to say chiral) quarks move at the speed of light.  Finally as mentioned earlier, the model lacks degeneracies between states with the same quantum numbers expected of a string theory.

To summarize, while the simple stringy model of ref.~\cite{A} {\it describes} the data in Fig.~\ref{stringfig} it is highly questionable as to whether it explains it.  One important open question is the nature of the correct stringy description in large $N_c$ QCD for highly-excited mesons including the spins of the quarks.

This leaves the situation somewhat muddled.  On reflection the analysis of ref. \cite{A} does highlight an important difficulty for the hypothesis of effective chiral restoration. However, the difficulty is empirical rather than theoretical---namely, the lack of parity partners for the highest spin states in the apparent bands.  It seems to me that  there are three likely resolutions to this: (i) The problem is with data. That is, with increased study the ``missing'' levels will be found restoring the agreement between the data and the conjecture of effective chiral restoration. As often noted earlier, the absence of evidence should not be taken as the evidence of absence.  Moreover, as {\it all} of the very highly excited states come from one type of reactions (proton-antiproton collisions)\cite{BUGG1,BUGG2,BUGG3,BUGG4,BUGG5}, it is possible that there are kinematical reasons why these missing states poorly couple to the incident channel.  It is noteworthy that the high lying  ``missing'' states all require higher partial waves then the observed states and this combined with phase space effects could suppress the production of these states in this kinematic region where the experiments were done\cite{Gcomment}.  Clearly there is a need to search for these states using experimental probes with different kinematics. (ii) The notion of effective chiral restoration might need to be modified somewhat.  It may happen that the onset of effect restoration might depend not merely on the excitation energy but also on the spin.  Thus the criterion may not be that the excitation energy is high, but that it is high compared to the lowest mass state of fixed $J$.  This will clearly help bring the data back into line with the conjecture of effective restoration.  On the other hand, it is by no means clear theoretically how to justify such behavior.  (iii) The conjecture is wrong.  It will be interesting to see which of these turns out to be right.

.
\end{document}